\documentclass[10pt,journal]{IEEEtran}
\UseRawInputEncoding
\usepackage[T1]{fontenc} % optional
\usepackage{setspace} % set space
\usepackage
[bookmarks=true,
colorlinks=true,
linkcolor=black,
anchorcolor=black,
urlcolor=black,
citecolor=black]{hyperref}
\usepackage{cite}
\usepackage{float}
\usepackage{graphicx}
\usepackage{caption}
\usepackage{subfigure}
\usepackage{tabularx}
\usepackage{booktabs}
\usepackage[table]{xcolor}
\newcolumntype{L}{>{\hspace*{-\tabcolsep}}l}
\newcolumntype{R}{c<{\hspace*{-\tabcolsep}}}
\definecolor{lightblue}{rgb}{0.93,0.95,1.0}
\usepackage{ifthen}
\usepackage{amsmath}
\usepackage{amssymb}
\usepackage{amsfonts}
\usepackage{commath}
\allowdisplaybreaks[4]

\newtheorem{mytheorem}{Theorem}

\newcommand{\rB}{\mathrm{B}}

\newcommand{\rn}{\mathrm{n}}

\newcommand{\rank}{\mathrm{rank}}
\newcommand{\trace}{\mathrm{tr}}

\newcommand{\LoS}{\mathrm{LoS}}
\newcommand{\NLoS}{\mathrm{NLoS}}

\newcommand{\SINR}{\mathrm{SINR}}

\newcommand{\USINR}{\underline{\SINR}}

\newcommand{\aslnr}{\mathrm{aslnr}}

\newcommand{\init}{\mathrm{init}}
\newcommand{\vmmse}{\mathrm{vmmse}}
\newcommand{\mmse}{\mathrm{mmse}}
\newcommand{\MMSE}{\mathrm{MMSE}}
\newcommand{\MSE}{\mathrm{MSE}}
\newcommand{\VMSE}{\mathrm{VMSE}}
\newcommand{\VMMSE}{\mathrm{VMMSE}}

\newcommand{\cps}{\mathrm{cps}}

\newcommand{\smallabs}[1]{\lvert#1\rvert}
\newcommand{\smallnorm}[1]{\lVert#1\rVert}

\newcommand{\dint}{\,\mathrm{d}}
\newcommand{\ub}{\mathrm{ub}}

\newcommand{\ut}{\mathrm{ut}}
\newcommand{\sat}{\mathrm{sat}}
\newcommand{\rx}{\mathrm{x}}
\newcommand{\ry}{\mathrm{y}}

\newcommand{\rv}{\mathrm{v}}

\newcommand{\Mx}{M_{\rx}}
\newcommand{\My}{M_{\ry}}

\newcommand{\Nx}{N_{\rx'}}
\newcommand{\Ny}{N_{\ry'}}
\newcommand{\Niter}{N_{\mathrm{iter}}}

\newcommand{\Nsc}{N_{\mathrm{sc}}}
\newcommand{\Ncp}{N_{\mathrm{cp}}}
\newcommand{\Tsc}{T_{\mathrm{sc}}}
\newcommand{\Tcp}{T_{\mathrm{cp}}}
\newcommand{\Ts}{T_{\mathrm{s}}}
\newcommand{\comma}{\text{,}}
\newcommand{\opt}{\mathrm{opt}}

\newcommand{\Complex}[2]{\bbC^{#1 \times #2}}
\newcommand{\Real}[2]{\bbR^{#1 \times #2}}
\newcommand{\xdeg}[1]{#1\text{\textdegree}}
\newcommand{\XH}[1]{\left(#1\right)^H}
\newcommand{\Xinv}[1]{\left(#1\right)^{-1}}
\newcommand{\xinv}[1]{\frac{1}{#1}}

\newcommand{\xiter}[2]{#1^{(#2)}}
\newcommand{\bbR}{\mathbb{R}}
\newcommand{\bbC}{\mathbb{C}}
\newcommand{\bbE}{\mathbb{E}}
\newcommand{\clF}{\mathcal{F}}

\newcommand{\clH}{\mathcal{H}}
\newcommand{\clI}{\mathcal{I}}
\newcommand{\clK}{\mathcal{K}}
\newcommand{\clL}{\mathcal{L}}
\newcommand{\clM}{\mathcal{M}}

\newcommand{\clR}{\mathcal{R}}

\newcommand{\clP}{\mathcal{P}}

\newcommand{\clS}{\mathcal{S}}

\newcommand{\clCN}{\mathcal{CN}}
\newcommand{\vphi}{\varphi}
\newcommand{\vtheta}{\vartheta}

\newcommand{\cklambda}{\check{\lambda}}
\newcommand{\ckbdH}{\check{\bdH}}

\newcommand{\htbdlambda}{\hat{\bdlambda}}

\newcommand{\htd}{\hat{d}}
\newcommand{\tdbeta}{\tilde{\beta}}
\newcommand{\tdlambda}{\tilde{\lambda}}

\newcommand{\tdtheta}{\tilde{\theta}}

\newcommand{\tdbdd}{\tilde{\bdd}}

\newcommand{\tdbdtheta}{\tilde{\bdtheta}}

\newcommand{\bdzro}{\mathbf{0}}
\newcommand{\bdone}{\mathbf{1}}
\newcommand{\bda}{\mathbf{a}}

\newcommand{\bdc}{\mathbf{c}}
\newcommand{\bdd}{\mathbf{d}}

\newcommand{\bdg}{\mathbf{g}}

\newcommand{\bdm}{\mathbf{m}}

\newcommand{\bds}{\mathbf{s}}

\newcommand{\bdu}{\mathbf{u}}
\newcommand{\bdv}{\mathbf{v}}
\newcommand{\bdw}{\mathbf{w}}
\newcommand{\bdx}{\mathbf{x}}
\newcommand{\bdy}{\mathbf{y}}
\newcommand{\bdz}{\mathbf{z}}
\newcommand{\bdA}{\mathbf{A}}
\newcommand{\bdB}{\mathbf{B}}
\newcommand{\bdC}{\mathbf{C}}

\newcommand{\bdH}{\mathbf{H}}
\newcommand{\bdI}{\mathbf{I}}

\newcommand{\bdQ}{\mathbf{Q}}
\newcommand{\bdR}{\mathbf{R}}

\newcommand{\bdT}{\mathbf{T}}

\newcommand{\bdV}{\mathbf{V}}
\newcommand{\bdW}{\mathbf{W}}

\newcommand{\bdomega}{\boldsymbol{\omega}}

\newcommand{\bdtheta}{\boldsymbol{\theta}}

\newcommand{\bdSigma}{\boldsymbol{\Sigma}}

\newcommand{\bdPhi}{\boldsymbol{\Phi}}

\newcommand{\bdlambda}{\boldsymbol{\lambda}}

\newcommand{\bdvphi}{\boldsymbol{\vphi}}
\newcommand{\stackeq}[1]{\stackrel{\text{(#1)}}{=}}
\newcommand{\stackleq}[1]{\stackrel{\text{(#1)}}{\le}}
\newcommand{\stackgeq}[1]{\stackrel{\text{(#1)}}{\ge}}

\usepackage{indentfirst}
\usepackage{algorithm,algorithmic}
\usepackage{cleveref}
\crefname{equation}{}{}
\crefname{figure}{Fig.}{Figs.}
\crefname{table}{TABLE}{TABLE}
\crefname{myprop}{Proposition}{Propositions}
\crefname{mycorollary}{Corollary}{Corollarys}
\crefname{mylemma}{Lemma}{Lemmas}
\crefname{mytheorem}{Theorem}{Theorems}
\Crefname{secinapp}{Appendix}{Appendices}

\begin{document}

\title{Downlink Transmit Design for Massive MIMO LEO Satellite Communications}

\author{Ke-Xin~Li,~\IEEEmembership{Student Member,~IEEE,}
	Li~You,~\IEEEmembership{Member,~IEEE,}
	Jiaheng~Wang,~\IEEEmembership{Senior Member,~IEEE,}\\
	Xiqi~Gao,~\IEEEmembership{Fellow,~IEEE,}
	Christos~G.~Tsinos,~\IEEEmembership{Senior Member,~IEEE,}\\
	Symeon~Chatzinotas,~\IEEEmembership{Senior Member,~IEEE,}
	and~Bj\"{o}rn~Ottersten,~\IEEEmembership{Fellow,~IEEE} % <-this % stops a space

\thanks{K.-X. Li, L. You, J. Wang and X. Q. Gao are with the National Mobile Communications Research Laboratory, Southeast University, Nanjing 210096, China (e-mail: likexin3488@seu.edu.cn; lyou@seu.edu.cn; jhwang@seu.edu.cn; xqgao@seu.edu.cn).}

\thanks{C. G. Tsinos, S. Chatzinotas and B. Ottersten are with the Interdisciplinary Centre for Security, Reliability and Trust (SnT), University of Luxembourg, Luxembourg City 2721, Luxembourg (e-mail: chtsinos@gmail.com; symeon.chatzinotas@uni.lu; bjorn.ottersten@uni.lu).}
}% <-this 

\maketitle

\begin{abstract}
	This paper investigates the downlink (DL) transmit design for massive multiple-input multiple-output (MIMO) low-earth-orbit (LEO) satellite communication systems, where only the slow-varying statistical channel state information is exploited at the transmitter. The channel model for the DL massive MIMO LEO satellite system is established, in which both the satellite and the user terminals (UTs) are equipped with uniform planar arrays. Observing the rank-one property of the channel matrices, we show that the single-stream precoding for each UT is the optimal choice that maximizes the ergodic sum rate. This favorable result simplifies the complicated design of transmit covariance matrices into that of precoding vectors without any loss of optimality. Then, an efficient algorithm is devised to compute the precoding vectors. Furthermore, we formulate an approximate transmit design based on the upper bound on the ergodic sum rate, for which the optimality of single-stream precoding still holds. We show that, in this case, the design of precoding vectors can be simplified into that of scalar variables, for which an effective algorithm is developed. In addition, a low-complexity learning framework is proposed for optimizing the scalar variables. Simulation results demonstrate that the proposed approaches can achieve significant performance gains over the existing schemes.
\end{abstract}


\IEEEpeerreviewmaketitle

%\newpage

%\tableofcontents
%\newpage
\section{Introduction}
\IEEEPARstart{E}{ver} increasing data demands present highly challenging requirements for future wireless networks, which are expected to provide extremely high throughputs, global and seamless coverage, ultra reliability, low latency and massive connectivity \cite{Guidotti2019Architecture5GSatellite}. 
As a critical enabler to achieve this ambitious target, satellite communication (SATCOM) can provide continuous and ubiquitous connectivity for areas without adequate Internet access \cite{3GPP_NonTerrestrial}.
In recent years, low-earth-orbit (LEO) satellites, typically deployed between $500$ km and $2000$ km from the earth, have attracted intensive research interest due to shorter round-trip delay, reduced pathloss and lower launch costs, compared to the geostationary-earth-orbit (GEO) satellites \cite{Qu2017LEOSatConste,Guidotti2019LTEbasedLEO,Di2019UltraDenseLEO,Su2019BroadbandLEO}.
Up to now, several projects have started by governments and corporations to develop LEO SATCOM systems, e.g., Iridium, Globalstar, OneWeb, Starlink, Telesat \cite{Fossa1998Iridium,Metzen2000Globalstar,Portillo2019Atechnical}.

Multibeam satellites, which serve a number of user terminals (UTs) on ground with spot beams, play an important role in SATCOM \cite{Maral2009SatelliteCommunications}.
Basically, the spot beams can be generated by using multifeed reflector antennas or phased-array antennas (PAAs) at the satellite side \cite{Lutz2000SatSysPerson}. 
While the GEO satellites are usually equipped with the multifeed reflector antennas \cite{Schneider2011Antenna}, the PAAs are more adapted for the LEO satellites because of their wide-angle coverage capabilities \cite{Lutz2000SatSysPerson}, e.g., Globalstar \cite{Metzen2000Globalstar} and Starlink \cite{Portillo2019Atechnical}.
In current satellite systems, multiple color reuse scheme is often adopted to suppress the inter-beam interference by exploiting different frequency bands and orthogonal polarizations \cite{Fenech2016Eutelsat}. In this case, the frequency bands have to be reused among sufficiently isolated beams to guarantee sufficient system capacity.

To exploit the limited spectrum more aggressively, full frequency reuse (FFR) scheme has been proposed, in which all beams share the same frequency band \cite{Vazquez2016PrecodingChallenges,PerezNeira2019SPforHTS}, thus improving the spectral efficiency. In this case, advanced signal processing techniques are indispensable to mitigate inter-beam interference.
To this end, precoding techniques arising from multiuser multiple-input multiple-output (MIMO) communications have been extensively studied in multibeam satellite systems \cite{Zheng2012GenericOptimization,Chrisopulos2015MultigroupFBSC,Joroughi2016PrecodingMultiGateway,Wang2018RobustMultigroup,Schwarz2019MIMOApplication}.
A generic precoding approach for a class of objective functions and power constraints was presented in \cite{Zheng2012GenericOptimization} for multibeam satellite systems. Based on the superframe structure in the DVB-S2X standard, the multi-group multicasting principle has been incorporated in the precoding for frame-based multibeam satellites \cite{Chrisopulos2015MultigroupFBSC,Joroughi2016PrecodingMultiGateway,Wang2018RobustMultigroup}. The distributed precoding for multi-gateway multibeam satellites can be found in \cite{Joroughi2016PrecodingMultiGateway}. 
In \cite{Schwarz2019MIMOApplication}, the antenna geometry in the MIMO feeder link and the zero-forcing (ZF) precoding in the multibeam downlink (DL) were studied.  

The previous works on the precoding for multibeam satellites generally assume that the beamforming network (BFN) at the satellite side is fixed \cite{Zheng2012GenericOptimization,Chrisopulos2015MultigroupFBSC,Joroughi2016PrecodingMultiGateway,Wang2018RobustMultigroup,Schwarz2019MIMOApplication}.
Indeed, the conventional BFN can only be modified in a very slow pace \cite{PerezNeira2019SPforHTS}, and unable to adapt to the link conditions of UTs timely. Nowadays, massive MIMO transmission has been widely accepted as one of the supporting techniques in terrestrial 5G communications \cite{Marzetta2010Noncooperative}.
By using a large number of antennas at the base station (BS), massive MIMO can provide substantial
degrees of freedom in the spatial domain, thus significantly improving the spectrum and energy efficiency \cite{Hien2013ESEfficiency}.
Essentially, the benefits of massive MIMO come from the multiple reconfigurable fine-grained beams, each of which is aligned to a specific UT. 
As the rapid development of 5G communications, a more flexible and versatile BFN can be digitally implemented at the satellite \cite{Hong2017MultibeamAntenna5G}, which can cater to the dynamic link conditions of UTs. 
In this paper, we focus on an LEO satellite system equipped with a massive antenna array, namely a massive MIMO LEO satellite, and we assume that the BFN at the LEO satellite can be digitally reconfigurable in real time, which is expected to enhance the throughput in wideband LEO SATCOM systems.

It is well known that the performance of multiuser MIMO/massive MIMO precoding critically depends on the quality of the channel state information at the transmitter (CSIT).
Most of the aforementioned works on the precoding in multibeam SATCOM systems assume that the transmitter can track the instantaneous CSI (iCSI) \cite{Zheng2012GenericOptimization,Chrisopulos2015MultigroupFBSC,Joroughi2016PrecodingMultiGateway,Schwarz2019MIMOApplication}. 
However, in practical SATCOM systems, the intrinsic channel impairments, e.g., large propagation delays and Doppler effects, will render it challenging to acquire the iCSIT. Particularly, for time-division duplexing (TDD) systems, the estimated uplink (UL) iCSI is used for the DL transmission, which may be outdated after the DL signals arrive at ground UTs. 
Meanwhile, in frequency-division duplexing (FDD) systems, the DL iCSI is first estimated at each UT and then fed back to the satellite, which
could bring considerable channel estimation and feedback overhead. Moreover, the feedback would also be outdated due to the large delays in SATCOM. In contrast to the iCSI, statistical CSI (sCSI) is valid for longer time intervals \cite{Gao2009StatisticalEigenmode}, and thus can be more easily obtained at the transmitter side. Hence, in this paper, we consider a practical scenario where only sCSI is available at the satellite to perform the DL transmit design in massive MIMO SATCOM. Here, we focus on the LEO satellites, although the presented design can be extended to the GEO ones.

The DL transmit design using sCSIT has received increasing attention in massive MIMO terrestrial wireless communications. Up to now, many transmit strategies have been presented, e.g., the two-stage precoder design \cite{AdhikaryJ2013SDM}, the beam domain transmission \cite{CSun2015BDMA}, and the robust precoder design \cite{AnLu2017RobustTransmission}.
However, the aforementioned works do not take the special massive MIMO LEO satellite channel characteristics into account. Also, the limited satellite payloads impose severe computational restrictions on the transmit design. Thus, it is imperative to seek out more efficient solutions for the DL transmit design with sCSIT in massive MIMO LEO SATCOM systems.

Recently, a transmission approach for massive MIMO LEO SATCOM systems was introduced in \cite{You2019MassiveMIMOLEO}, where the channel model, the DL precoders and UL receivers, and the user grouping strategy were investigated. 
Note that, in \cite{You2019MassiveMIMOLEO}, each UT only has a single antenna, thus restricting the performance of the massive MIMO LEO SATCOM system. 
Moreover, the DL precoding vectors in \cite{You2019MassiveMIMOLEO} are based on an individual performance metric called the average signal-to-leakage-plus-noise ratio (ASLNR), and not on one that captures the overall performance of the whole system. Therefore, the transmission scheme proposed in \cite{You2019MassiveMIMOLEO} is insufficient to fully exploit the potential of massive MIMO technique in LEO SATCOM systems. In \cite{Angeletti2020MassiveMIMOSatellite}, the switching-based beam selection scheme and radio resource management strategy were jointly considered, in which only the location information of UTs is exploited to assign the beams from a set of fixed beams. Nevertheless, this scheme does not make full use of the flexible beamforming capabilities of massive MIMO, which leads to a certain degree of degradation in system performance.

In this paper, we consider the massive MIMO LEO SATCOM system where the satellite and the UTs are both equipped with uniform planar arrays (UPAs). We investigate how to achieve high data rates of the whole system using only the slow-varying sCSIT to properly design the DL transmit strategy. For this purpose, we first derive the DL massive MIMO LEO satellite channel model with the UPA configurations at the satellite and each UT. The adverse Doppler and delay effects are compensated by performing frequency and time synchronization at each UT to facilitate the DL wideband transmission. Then, based on the massive MIMO LEO satellite channel characteristics, we propose the DL transmit design, and aim to maximize the ergodic sum rate of all UTs by exploiting sCSIT. Our major contributions are summarized as follows

\begin{itemize}
	\item We find that the single-stream transmit strategy for each UT is optimal for the linear transmitters in the sense of maximizing the system's ergodic sum rate, even though each UT has multiple antennas.
	This result is important and favorable because the complicated design of transmit covariance matrices can be simplified into that of precoding vectors without any loss of optimality.
	Then, we devise an algorithm to compute the precoding vectors.
	\item To reduce the computational complexity, we formulate another transmit design by approximating the ergodic sum rate with its upper bound. In this case, it is shown that the optimality of the single-stream transmit strategy still holds. More importantly, the design of precoding vectors is further simplified to that of scalar variables, for which an algorithm is developed.
	\item In addition, a learning-based solution is proposed for the scalar-variable optimization problem with significantly reduced onboard implementation complexity. 
	Simulation results demonstrate the effectiveness of the proposed approaches, and show remarkable performance gains over the existing schemes.
\end{itemize}

The remainder of this paper is organized as follows. \Cref{Section_system_model} introduces the system model, where the channel model is presented for the satellite and the UTs equipped with UPAs. In \Cref{Section_DL_Precoder_Design}, the rank-one property of the transmit covariance matrices is proved and the precoding algorithm by considering the ergodic sum rate maximization is presented. In \Cref{Section_Low_Complexity_DL_Precoder_Design}, we present another transmit design with the upper bound on the ergodic sum rate, and a low-complexity learning-based approach is proposed. \Cref{Sectioin_Simulation} provides the simulation results, and \Cref{Section_Conclusion} concludes this paper.

\textit{Notations:} Throughout this paper, lower case letters denote scalars, and boldface lower (upper) letters denote vectors (matrices). The set of all $n$-by-$m$ complex (real) matrices is denoted as $\bbC^{n\times m}$ ($\bbR^{n\times m}$). $\trace(\cdot)$, $\det(\cdot)$, $\rank(\cdot)$, $(\cdot)^*$, $(\cdot)^T$, and $(\cdot)^H$ denote the trace, determinant, rank, conjugate, transpose, and conjugate transpose operations for the matrix argument, respectively. $\abs{\cdot}$ denotes the absolute value. The Euclidean norm of a vector $\bdx$ is denoted as $\norm{\bdx} = \sqrt{\bdx^H \bdx}$. $\otimes$ denotes the Kronecker product. $\clCN(\bdzro,\bdC)$ denotes the circular symmetric complex Gaussian random vector with zero mean and covariance matrix $\bdC$. 
%$\mathrm{U}\ [a,b]$ represents the uniform distribution between $a$ and $b$. 
%$\triangleq$ denotes ``be defined as''. $\sim$ denotes ``be distributed as''.

\section{System Model}	\label{Section_system_model}

\subsection{System Setup}	\label{subsec_system_setup}
\begin{figure}[!t]
	\centering
	\vspace{-0em}
	\includegraphics[width=0.45\textwidth]{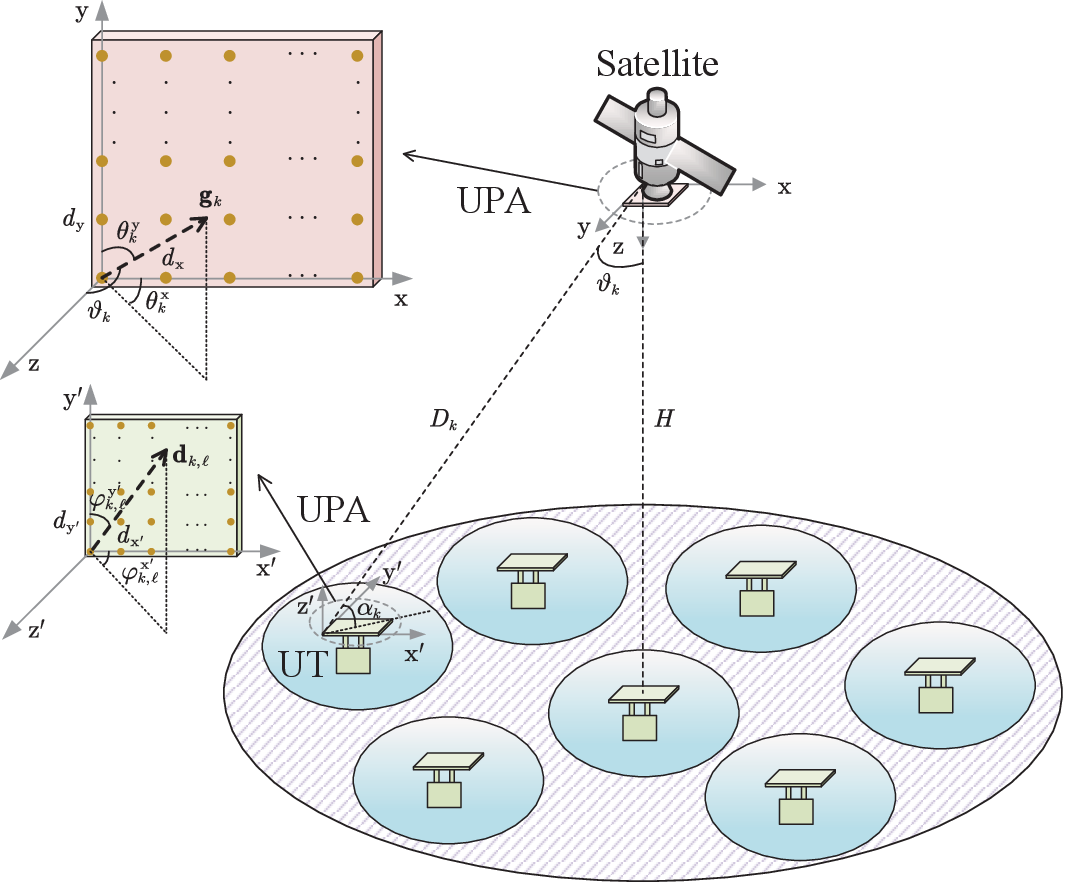}
	\caption{The DL in FFR massive MIMO LEO SATCOM.}
	\label{fig_UPA}
	\vspace{-0em}
\end{figure}

We consider the DL transmission in an FFR massive MIMO LEO SATCOM system over lower frequency bands, e.g., L/S/C bands.
The mobile UTs are served by a single LEO satellite at an altitude of $H$ as shown in \Cref{fig_UPA}.
The satellite is assumed to work with a regenerative payload, which allows on-board processing (OBP) of baseband signals on the satellite \cite{PerezNeira2019SPforHTS}.
The satellite and the mobile UTs are equipped with the UPAs of digital active antennas \cite{Hong2017MultibeamAntenna5G}, which means that the amplitude and phase on each antenna element of the UPAs can be digitally controlled. 
The satellite has a large-scale UPA with $\Mx$ and $\My$ elements in the $\rx$-axis and $\ry$-axis, respectively. The total number of antennas at the satellite is $\Mx \My \triangleq M$. 
We assume that each antenna element of the UPA at the satellite is directional.
On the other hand, each UT's UPA consists of $\Nx$ and $\Ny$ omnidirectional elements in the $\rx'$-axis and $\ry'$-axis, respectively, and the total number of antennas at each UT is $\Nx \Ny \triangleq N$. The approach in this paper can be directly extended to the case where the UPAs at UTs have different numbers of antenna elements. 

\subsection{Signal and Channel Models in Analog Baseband} \label{subsec_DL_channel_model}
The DL received signal at UT $k$ at the time instant $t$ can be written as 
\begin{equation}
	\bdy_{k}(t) = \int_{-\infty}^{\infty} \ckbdH_{k}(t,\tau) \bdx(t-\tau) \dint \tau + \bdz_{k}(t)\comma
\end{equation}
where $\ckbdH_{k}(t,\tau) \in \Complex{N}{M}$, $\bdx(t) \in \Complex{M}{1}$ and $\bdz_{k}(t) \in \Complex{N}{1}$ are the channel impulse response, transmit signal and additive noise signal of UT $k$ at time instant $t$, respectively.
More specifically, the LEO satellite channel impulse response $\ckbdH_{k}(t,\tau)$ can be expressed as
\begin{equation}
	\ckbdH_{k}(t,\tau) = \sum_{\ell=0}^{L_k-1} a_{k,\ell}  e^{ j2\pi \nu_{k,\ell} t } \delta \left( \tau-\tau_{k,\ell} \right)  \bdd_{k,\ell} \bdg_{k,\ell}^H\comma \label{channel_model_DL_k}
\end{equation}
where $j \triangleq \sqrt{-1}$, $\delta(x)$ is the Dirac delta function, $L_k$ is the multipath number of UT $k$'s channel, $a_{k,\ell}$, $\nu_{k,\ell}$, $\tau_{k,\ell}$, $\bdd_{k,\ell} \in \Complex{N}{1}$ and  $\bdg_{k,\ell} \in \Complex{M}{1}$ are the complex channel gain, Doppler shift, propagation delay, array response vector at the UT side and array response vector at the satellite side, respectively, associated with the $\ell$th path of UT $k$'s channel. 
%Correspondingly, the channel frequency response $\ckbdH_k(t,f) \in \Complex{N}{M}$ is given by
%\begin{equation}
%
%\ckbdH_k(t,f) = \sum_{\ell=0}^{L_k-1} a_{k,\ell}  e^{ j2\pi \left( \nu_{k,\ell} t - f \tau_{k,\ell}  \right) } \bdd_{k,\ell} \cdot \bdg_{k,\ell}^H. 
%\end{equation} 

For simplicity, we assume that the channel matrices are fixed within each coherence time interval, and change from block to block according to some ergodic process. 
In the following, we will describe the LEO satellite channel characteristics one by one, which mainly include the Doppler shifts, propagation delays, and array response vectors.

\subsubsection{Doppler shifts} For LEO satellite channels, the Doppler shifts will be much larger compared with those in terrestrial wireless channels, due to the large relative velocity between the satellite and  the UTs. At the $4$ GHz carrier frequency, the Doppler shift can be $80$ kHz for an LEO satellite at an altitude of $1000$ km \cite{Ali1998DopplerLEO}.
The Doppler shift $\nu_{k,\ell}$ for the $\ell$th path of UT $k$'s channel mainly consists of two parts \cite{Papath2001Acomparison}, i.e.,
$\nu_{k,\ell} = \nu_{k,\ell}^{\sat} + \nu_{k,\ell}^{\ut}$,
where $\nu_{k,\ell}^{\sat}$ and $\nu_{k,\ell}^{\ut}$ are the Doppler shifts relevant to the movement of the satellite and UT $k$, respectively.	
The first part $\nu_{k,\ell}^{\sat}$ is nearly identical for different paths of UT $k$'s channel, because of the high altitude of the satellite \cite{Papath2001Acomparison}. Hence, $\nu_{k,\ell}^{\sat}$ can be rewritten as $\nu_{k,\ell}^{\sat} = \nu_{k}^{\sat}$ for $0\le \ell \le L_k-1$.  
The variation of $\nu_{k}^{\sat}$ with time behaves rather deterministically, and it can be estimated and compensated at each UT. 
Specifically, $\nu_k^{\sat}$ can be expressed as $\nu_k^{\sat} = f_c (v_k/c) \cos \phi_k$ \cite{Ali1998DopplerLEO},
where $f_c$ is the carrier frequency, $c$ is the speed of light, $v_k$ is the velocity of the satellite, and $\phi_k$ is the angle between the satellite's forward velocity and boresight from the satellite to UT $k$. 
On the other hand, the $\nu_{k,\ell}^{\ut}$'s are usually distinct for different paths. 
	
\subsubsection{Propagation Delays} For LEO satellites, the propagation delay is a more serious problem than that in terrestrial wireless channels, due to the long distance between the satellite and the UTs. For an LEO satellite at an altitude of $1000$ km, the round-trip delay is about $17.7$ ms with $\xdeg{45}$ of elevation angles \cite{Guidotti2019LTEbasedLEO}. We use $\tau_k^{\min} = \min_{\ell} \tau_{k,\ell}$ and $\tau_k^{\max} = \max_{\ell} \tau_{k,\ell}$ to represent the minimal and maximal propagation delays of UT $k$'s channel, respectively. 

\subsubsection{Array response vectors} Define $\bdtheta_{k,\ell} = ( \theta_{k,\ell}^{\rx},\theta_{k,\ell}^{\ry} )$ and $\bdvphi_{k,\ell} = ( \vphi_{k,\ell}^{\rx'},\vphi_{k,\ell}^{\ry'} )$ as the paired angles-of-departure (AoDs) and angles-of-arrival (AoAs) for the $\ell$th path of UT $k$'s channel, respectively. The array response vectors $\bdg_{k,\ell}$ and $\bdd_{k,\ell}$ in \eqref{channel_model_DL_k} are given by $\bdg_{k,\ell} = \bdg ( \bdtheta_{k,\ell} ) \label{g_k,l}$ and $\bdd_{k,\ell} = \bdd ( \bdvphi_{k,\ell} )$, respectively,
where $\bdg(\bdtheta) = \bda_{\Mx} \left( \sin \theta_{\ry} \cos \theta_{\rx} \right) \otimes 
\bda_{\My} \left( \cos \theta_{\ry} \right)$ and $\bdd(\bdvphi) =  \bda_{\Nx} \left( \sin \vphi_{\ry'} \cos \vphi_{\rx'} \right) \otimes \bda_{\Ny} \left( \cos \vphi_{\ry'} \right) $ for arbitrary $\bdtheta=(\theta_{\rx},\theta_{\ry})$ and $\bdvphi=(\vphi_{\rx'},\vphi_{\ry'})$.
Here, $\bda_{n_\rv}(x) \in \Complex{n_\rv}{1}$ is expressed as $\bda_{n_\rv} \left( x \right) = \frac{1}{\sqrt{n_\rv}} ( 1, e^{-j\frac{2\pi d_{\rv}}{\lambda} x }, \dots, e^{-j\frac{2\pi d_{\rv}}{\lambda} (n_\rv-1) x } )^T$, where $\lambda=c/f_c$ is the carrier wavelength, $d_{\rv}$ is the antenna spacing along $\rv$-axis with $\rv\in\{ \rx,\ry,\rx',\ry' \}$.
In satellite channels, the scattering on ground takes place only within a few kilometers around each UT. Thus, the paired AoDs for different paths of UT $k$'s channel are nearly identical due to the long distance between the satellite and UT $k$ \cite{You2019MassiveMIMOLEO}, i.e., $\bdtheta_{k,\ell} = \bdtheta_k$, $0 \le \ell \le L_k-1$. Therefore, we can rewrite $\bdg_{k,\ell} = \bdg_{k} = \bdg(\bdtheta_k)$,
where $\bdtheta_k = (\theta_k^{\rx},\theta_k^{\ry})$ is referred to as the physical angle pair of UT $k$. 
Due to the long distance between the satellite and UT $k$, $\bdg_k$ changes quite slowly, and we assume that it can be perfectly known at the satellite.
The space angle pair $\tdbdtheta_k = (\tdtheta_k^{\rx},\tdtheta_k^{\ry})$ of UT $k$ is defined as $\tdtheta_k^{\rx} = \sin \theta_k^{\ry} \cos \theta_k^{\rx} $ and $\tdtheta_k^{\ry} = \cos \theta_k^{\ry} $, which reflects the space domain property of UT $k$'s channel \cite{You2019MassiveMIMOLEO}. 
The physical angle pair $\bdtheta_k$ and nadir angle 
$\vtheta_k$ of UT $k$ as depicted in \Cref{fig_UPA} are related by $\cos \vtheta_k = \sin \theta_k^{\ry} \sin \theta_k^{\rx}$.

%To alleviate the influence of the large Doppler shifts and propagation delays in the LEO satellite channel, we will elaborate the Doppler and delay compensation techniques in \Cref{subsec_Doppler_delay_compensate}.

\subsection{Signal and Channel Models for OFDM Based Transmission} \label{subsec_Doppler_delay_compensate}
We consider that the orthogonal frequency division multiplex (OFDM) is used to facilitate the wideband transmission in the LEO SATCOM systems.
The number of subcarriers is $\Nsc$, and the cyclic prefix (CP) length is $\Ncp$. Let $\Ts$ be the system sampling period. The time duration of CP is $\Tcp = \Ncp \Ts$. The OFDM symbol time duration without and with CP is given by $\Tsc = \Nsc \Ts$ and $T =\Tsc + \Tcp$, respectively. 

Let $\{ \bdx_{s,r} \}_{r=0}^{\Nsc-1}$ be the $M\times 1$ frequency-domain transmit signal within the $s$th OFDM symbol. Then, the time-domain transmit signal
in OFDM symbol $s$ can be expressed as \cite{Hwang2009OFDMSurvey}
\begin{equation}
\bdx_{s}(t) = \sum_{r=0}^{\Nsc-1} \bdx_{s,r} e^{j2\pi r \Delta f t}\comma\ -\Tcp \le t-sT < \Tsc\comma
\end{equation}
where $\Delta f = 1/\Tsc$. The time-domain received signal of UT $k$ in the OFDM symbol $s$ can be written as
\begin{equation}
	\bdy_{k,s}(t) = \int_{-\infty}^{\infty} \ckbdH_{k}(t,\tau) \bdx_s (t-\tau) \dint \tau + \bdz_{k,s}(t)\comma
\end{equation}
where $\bdz_{k,s}(t)$ is the additive noise signal of UT $k$ at the OFDM symbol $s$.
Next, by exploiting the LEO satellite channel characteristics, joint Doppler and delay compensation is applied at each UT. Let $\nu_k^{\cps} = \nu_{k}^{\sat}$ and $
\tau_k^{\cps} = \tau_{k}^{\min}$. Based on the results in \cite{You2019MassiveMIMOLEO}, the compensated time-domain received signal of UT $k$ in the OFDM symbol $s$ is given by
\begin{equation}
\bdy_{k,s}^{\cps}(t) = \bdy_{k,s}(t+\tau_k^{\cps}) e^{-j2\pi \nu_k^{\cps} ( t + \tau_k^{\cps} ) }.
\end{equation}
After the Doppler and  delay compensation, we choose the well-designed OFDM parameters to combat the multipath fading effect. Hence, the frequency-domain received signal of UT $k$ over the subcarrier $r$ in the OFDM symbol $s$ can be written as \cite{Hwang2009OFDMSurvey}
\begin{equation}
\bdy_{k,s,r} = \xinv{\Tsc} \int_{sT}^{sT+\Tsc} \bdy_{k,s}^{\cps}(t) e^{-j2\pi r \Delta f \cdot t} \dint t. \label{yksr_digitl_domain}
\end{equation}
Let us denote $\tau_{k,\ell}^{\ut} = \tau_{k,\ell} - \tau_k^{\min}$, and define the effective channel frequency response of UT $k$ after the Doppler and delay compensation as
\begin{equation}
\bdH_{k}(t,f) = \bdd_k(t,f) \bdg_k^H\comma \label{channel_model_UTk_compensate}
\end{equation} 
where 
$\bdd_k(t,f) = \sum_{\ell=0}^{L_k-1} a_{k,\ell}  e^{ j 2\pi \left( \nu_{k,\ell}^{\ut} t - f \tau_{k,\ell}^{\ut} \right) } \bdd_{k,\ell} \in \Complex{N}{1}$.
Then, the received signal $\bdy_{k,s,r}$ in \eqref{yksr_digitl_domain} can be further expressed  as
\begin{equation}
	\bdy_{k,s,r} = \bdH_{k,s,r} \bdx_{s,r} + \bdz_{k,s,r}\comma \label{yksr_Hksr_xksr_zksr}
\end{equation}
where $\bdH_{k,s,r}$ and $\bdz_{k,s,r}$ are the channel matrix and additive Gaussian noise of UT $k$ over the subcarrier $r$ in the OFDM symbol $s$. 
Note that $\bdH_{k,s,r}$ in \eqref{yksr_Hksr_xksr_zksr} can be written as
\begin{equation}
	\bdH_{k,s,r} = \bdH_k \left( sT, r\Delta f \right) =  \bdd_{k,s,r}  \bdg_k^H\comma \label{Hksr_dksr_gk}
\end{equation}
where $\bdd_{k,s,r} = \bdd_k \left( sT, r\Delta f \right)$.
Since the Doppler and the delay effects are compensated at each UT, the time and frequency at the satellite and the UTs are assumed to be perfectly synchronized in the following. 

\subsection{Statistical Properties of Channel} \label{subsec_statistic_channel_model}
To describe the statistical properties of the channel matrices conveniently, we omit the subscripts of OFDM symbol $s$ and subcarrier $r$ in $\bdH_{k,s,r} = \bdd_{k,s,r} \bdg_k^H$ and denote $\bdH_k = \bdd_k \bdg_k^H$ as the DL channel matrix of UT $k$ over a specific subcarrier. 
In this paper, the channel $\bdH_k$ is supposed to be Rician distributed as follows
\begin{equation} 
	\bdH_k = \bdd_k \bdg_k^H = \sqrt{\frac{\kappa_k \beta_k}{\kappa_k + 1}} \bdH_{k}^{\LoS} + \sqrt{\frac{\beta_k}{\kappa_k + 1}} \bdH_{k}^{\NLoS}\comma \label{channel_matrix_UTk}
\end{equation}
where $\beta_k = \bbE \left\{ \trace(\bdH_k \bdH_k^H) \right\} = \bbE \left\{ \smallnorm{\bdd_k}^2 \right\}$ is the average channel power, $\kappa_k$ is the Rician factor, $\bdH_{k}^{\LoS} = \bdd_{k,0} \bdg_k^H$ is the deterministic line-of-sight (LoS) part, and $\bdH_{k}^{\NLoS} = \tdbdd_{k} \bdg_k^H$ is the random scattering part. Besides, $\tdbdd_{k}$ is distributed as $\tdbdd_{k} \sim \clCN(\bdzro,\bdSigma_k)$ with $\trace(\bdSigma_k) = 1$.
The channel parameters $\clH \triangleq \{ \beta_k, \kappa_k, \bdg_k,\bdd_{k,0}, \bdSigma_k \}_{\forall k}$ are related to the operating frequency bands, the practical link conditions, and so on \cite{Lutz2000SatSysPerson}. 
We also assume that the satellite and the UTs move within a certain range, such that the channel parameters $\clH$ can be considered as nearly unchanged. Whenever the satellite or some UT steps out of this range, the channel parameters $\clH$ should be updated at accordingly.

The channel correlation matrices of UT $k$ at the satellite and the UT sides are given by
\begin{subequations}
	\begin{align}
		\bdR_k^{\sat} &= \bbE \{ \bdH_k^H \bdH_k \} = \beta_k \bdg_k \bdg_k^H\comma \\
		\bdR_k^{\ut} &= \bbE \{ \bdH_k \bdH_k^H \} = \frac{\kappa_k \beta_k }{\kappa_k + 1} \bdd_{k,0} \bdd_{k,0}^H + \frac{\beta_k}{\kappa_k + 1} \bdSigma_{k}\comma
	\end{align}
\end{subequations}
respectively.
The matrix $\bdR_k^{\sat}$ is rank-one, which implies that the signals on different antennas at the satellite are highly correlated. Meanwhile, the rank of matrix $\bdR_k^{\ut}$ depends on the specific propagation environment around UT $k$.

\section{Transmit Design} \label{Section_DL_Precoder_Design}
In this section, we investigate the DL transmit design for the examined massive MIMO LEO SATCOM system based on the established signal and channel models in \Cref{Section_system_model}. First, by exploiting the LEO satellite channel characteristics, we prove that the rank of transmit covariance matrix of each UT must be no greater than one to maximize the ergodic sum rate. This indicates that the optimal DL transmission strategy is to transmit a single data stream to each UT, even if each UT has multiple antennas. This result is particularly important since the original design of transmit covariance matrices can be simplified into that of the precoding vectors without any loss of optimality. Based on this result, we develop an algorithm, by invoking the minorization-maximization (MM) framework, to efficiently compute the precoding vectors.

\subsection{Rank-One Property of Transmit Covariance Matrices} \label{subsec_DL_Transmission_Model}
By dropping the subscripts of OFDM symbol $s$ and subcarrier $r$ in $\bdx_{s,r}$ for simplicity, we denote $\bdx \in \Complex{M}{1}$ as the transmit signal at the satellite over a specific subcarrier. We consider that $K$ UTs are simultaneously served in the DL transmission.
The set of UT indices is denoted as $\clK = \left\{1,\dots,K\right\}$. 
The transmit signal $\bdx$ can be expressed as
\begin{equation}
	\bdx = \sum_{k=1}^K \bds_k\comma \label{x_sum_sk}
\end{equation} 
where $\mathbf{s}_k \in \Complex{M}{1}$ is the transmit signal related to UT $k$. In this paper, we consider the most general design of the transmit signals $\{\bds_k\}_{k=1}^K$, where $\bds_k$ is a Gaussian random vector with zero mean and covariance matrix $\bdQ_k = \bbE\{\bds_k \bds_k^H \}$.
For simplicity, we assume that the DL transmission satisfies a constraint on the total transmit power as in \cite{Zheng2012GenericOptimization,Joroughi2016PrecodingMultiGateway}, i.e.,  $\sum_{k=1}^{K} \trace(\bdQ_k) \le P$, although per-antenna power constraint may be more relevant for practice \cite{Dimitrios2015Multicast}.
The DL received signal at UT $k$ is given by
\begin{equation}
\bdy_k = \bdH_k \sum_{i=1}^{K} \bds_i + \bdz_k\comma \label{Transmission_Model_DL}
\end{equation}
where $\bdz_k \in \Complex{N}{1}$ is the additive complex Gaussian noise at UT $k$ distributed as $\bdz_k \sim \clCN \left(0,\sigma_k^2 \bdI_N \right)$.
The DL ergodic rate of UT $k$ is defined as
\begin{align}
\clI_k 
&= \mathbb{E} \left\{ \log \det \left( \sigma_k^2 \bdI_N + \bdH_k \sum_{i=1}^{K} \bdQ_i \bdH_k^H \right) \right\} \notag \\
&\qquad - \mathbb{E} \left\{ \log \det \left( \sigma_k^2 \bdI_N + \bdH_k \sum_{i \ne k} \bdQ_i \bdH_k^H \right)  \right\} \notag \\
&\stackeq{a} \mathbb{E} \left\{ \log \left( 1 + \frac{  \bdg_k^H \bdQ_k \bdg_k \norm{\bdd_k}^2 }{ \sum_{i \ne k} \bdg_k^H \bdQ_i \bdg_k \norm{\bdd_k}^2 + \sigma_k^2 } \right) \right\}\comma \label{DL_ergodic_rate_k_noRx}
\end{align}
where (a) follows from $\bdH_k = \bdd_k \bdg_k^H$ and $\det(\bdI + \bdA \bdB) = \det(\bdI+\bdB\bdA)$ \cite{Horn2013MatrixAnalysis}. 
The DL sum rate maximization problem can be formulated as\footnote{The weight factors can be introduced straightforwardly to consider the priorities of UTs.}
\begin{subequations}\label{Problem_SumRate_Max_Covariance}
\begin{align}
\clP: \ \max_{ \left\{ \bdQ_k \right\}_{k=1}^K }\ & \sum_{k=1}^{K} \clI_k\\ 
\mathrm{s.t.}\ & \sum_{k=1}^{K} \trace(\bdQ_k) \le P, \ \bdQ_k \succeq \bdzro, \ \forall k \in \clK. 
\end{align}
\end{subequations}

\begin{mytheorem} \label{Prop_rank-one_Covariance}
	The optimal $\{\bdQ_k\}_{k=1}^K$ to problem $\clP$ must satisfy $\rank(\bdQ_k) \le 1$, $\forall k \in \clK$.
\end{mytheorem}
\begin{IEEEproof}
	Please refer to \Cref{appendix_rank-one_Covariance_proof}.
\end{IEEEproof}

In \Cref{Prop_rank-one_Covariance}, we show that the rank of the optimal transmit covariance matrix of each UT should be no larger than one. Since $\rank(\bdQ_k)$ represents the number of independent data streams transmitted to UT $k$, \Cref{Prop_rank-one_Covariance} reveals that the single-stream precoding strategy for each UT is optimal for linear transmitters even though each UT has multiple antennas.
Following the rank-one property of the transmit covariance matrices, we express $\bdQ_k$ as $\bdQ_k = \bdw_k \bdw_k^H$, where $\bdw_k \in \Complex{M}{1} $ is the precoding vector of UT $k$.
Since $\{\bdw_k\}_{k=1}^K$ denote the linear precoding vectors, the transmit signal $\bds_k $ in \eqref{x_sum_sk} is expressed as $\bds_k = \bdw_k s_k$, where $s_k$ is the desired data symbol for UT $k$ with zero mean and unit variance. 
Henceforth, the design of the transmit covariance matrices $\{
\bdQ_k\}_{k=1}^K$ is now simplified into that of the precoding vectors $\{\bdw_k\}_{k=1}^K$.
Substituting $\bdQ_k = \bdw_k \bdw_k^H$ into \eqref{DL_ergodic_rate_k_noRx} yields
\begin{equation}
\clI_k = \mathbb{E} \left\{ \log \left( 1 + \frac{  \abs{\bdw_k^H \bdg_k}^2 \norm{\bdd_k}^2 }{ \sum_{i \ne k} \abs{\bdw_i^H \bdg_k}^2 \norm{\bdd_k}^2 + \sigma_k^2 } \right) \right\} \triangleq \clR_k. \label{DL_ergodic_rate_UT_k_precoder}
\end{equation}
Here, we replace $\clI_k$ with $\clR_k$ to represent the DL ergodic rate of UT $k$, since $\clR_k$ is now a function of the linear precoding vectors $\{ \bdw_k \}_{k=1}^K$.
Thus, the complicated transmit covariance matrix optimization problem $\clP$ in \eqref{Problem_SumRate_Max_Covariance} can be reformulated as follows
\begin{equation}
\clS: \ \max_{ \bdW  } \ \sum_{k=1}^{K} \clR_k\comma \quad \mathrm{s.t.} \ \sum_{k=1}^{K} \lVert \mathbf{w}_k \rVert^2 \le P\comma	\label{Problem_DL_Tx_design_Inst_Rx}
\end{equation}
where $\bdW = [ \bdw_1 \ \cdots \ \bdw_K ] \in \bbC^{M \times K}$ denotes the collection of the precoding vectors.
The power inequality in \eqref{Problem_DL_Tx_design_Inst_Rx} must be met with equality at the optimum, i.e., $\sum_{k=1}^{K} \lVert \mathbf{w}_k \rVert^2 = P$. Otherwise, $\{\bdw_k\}_{k=1}^K$ can be scaled up, which increases the DL sum rate and contradicts the optimality. 

Although we focus on the DL transmit design in this paper, the optimal linear receivers at the UT sides are also obtained as the by-product. In the following subsection, we derive the optimal linear receivers that maximize their corresponding DL ergodic rates.

\subsection{Optimal Linear Receivers}
According to \Cref{Prop_rank-one_Covariance}, the satellite can send at most one data stream to each UT. Hence, each UT just needs to decode at most one data stream, and only diversity gain is obtained with multiple antennas at the UT sides. 
Let $\mathbf{c}_k \in \Complex{N}{1}$ be the linear receiver of UT $k$. 
Then, the recovered data symbol at UT $k$ can be written as
\begin{align}
\hat{s}_k 
&= \bdc_k^H \bdy_k  \notag \\
&= \bdc_k^H \bdd_k \bdg_k^H \bdw_k s_k + \sum_{i \ne k}^K \bdc_k^H \bdd_k \bdg_k^H \bdw_i s_i + \bdc_k^H \bdz_k. \label{hat_sk}
\end{align}
Thus, the signal-to-interference-plus-noise ratio (SINR) of UT $k$ can be expressed as
\begin{equation}
\SINR_k
= \frac{ \left\lvert \bdw_k^H \bdg_k \right\rvert^2 \left\lvert \bdc_k^H \bdd_k \right\rvert^2 }{ \sum_{i \ne k} \left\lvert \bdw_i^H \bdg_k \right\rvert^2 \left\lvert \bdc_k^H \bdd_k \right\rvert^2  + \sigma_k^2 \left\lVert \bdc_k \right\rVert^2 }.
\end{equation}
Because $\frac{ax}{bx+c}$ is a monotonically increasing function of $x$ for $a,b,c > 0$,  we have
\begin{equation}
\SINR_k 
\stackleq{a} \frac{ \left\lvert \bdw_k^H \bdg_k \right\rvert^2 \norm{\bdd_k}^2 }{ \sum_{i \ne k} \left\lvert \bdw_i^H \bdg_k \right\rvert^2 \norm{\bdd_k}^2  + \sigma_k^2 } \triangleq \USINR_k\comma \label{DL_SINR_upperbound_iRx}
\end{equation}
where (a) follows from the Cauchy-Schwarz inequality $\smallabs{\bdc_k^H \bdd_k}^2 \le \norm{\bdc_k}^2 \norm{\bdd_k}^2$, and the equality holds if and only if $\bdc_k = \alpha \bdd_k$ for any nonzero $\alpha \in \bbC$. 
The receivers satisfying $\bdc_k = \alpha \bdd_k$ for different $\alpha$ will have the same value of $\SINR_k$.
Thus, the receivers with the form $\bdc_k = \alpha \bdd_k$ are optimal for UT $k$.
Now, we will return to the precoding vector design in the following subsection.

\subsection{Precoding Vector Design}
In this subsection, we aim to compute the precoding vectors by maximizing the ergodic sum rate under the discussed sum power constraint. Considering that the precoding vector optimization problem $\clS$ in \eqref{Problem_DL_Tx_design_Inst_Rx} is a non-convex program, we develop an algorithm based on the MM framework \cite{DavidTutorialMM} to compute the precoding vectors. 

In the following, we develop an MM-based algorithm to obtain a locally optimal solution to $\clS$. In each iteration, the DL ergodic rate $\clR_k$ is replaced with its concave minorizing function. Then, a locally optimal solution to $\clS$ can be obtained by iteratively solving a sequence of convex programs. By making use of the relationship between the ergodic rate and the minimum mean-square error (MMSE) \cite{AnLu2017RobustTransmission}, for given precoders $\bdW^{(n)}$ in the $n$th iteration, we can derive a minorizing function of $\clR_k$ as 
\begin{align}
\xiter{g_k}{n} &= - \left( a_k^{(n)} \sum_{i=1}^K \left\lvert \bdw_i^H \bdg_k \right\rvert^2 - 2 \Re \left\{ \bdw_k^H \bdg_k \cdot b_k^{(n)} \right\}\right)\notag \\
&\qquad   - c_k^{(n)} + 1 + \xiter{\clR_k}{n}\comma \label{Minorize_udR_DL_InstRx}
\end{align}
where $a_k^{(n)}$, $b_k^{(n)}$ and $c_k^{(n)}$ are constants defined in \Cref{appendix_minorize_udR_DL_instRx_proof}.
By using the minorizing function $\xiter{g_k}{n}$ in \eqref{Minorize_udR_DL_InstRx}, the precoders $\bdW^{(n+1)}$ in the $(n+1)$th iteration can be obtained by solving the following convex program
\begin{equation}
\clS^{(n)}:\ \max_{\bdW} \ \sum_{k=1}^{K} \xiter{g_k}{n}\comma \quad
\mathrm{s.t.} \ \sum_{k=1}^{K} \norm{ \bdw_k }^2 \le P\comma \label{Problem_DL_Tx_design_Inst_Rx_n+1_Rate}
\end{equation}
which is equivalent to
\begin{subequations}
\begin{align}
	\clS^{(n)}: \ \min_{\bdW}\ & \sum_{k=1}^{K} \left( \sum_{i=1}^{K} a_i^{(n)} \abs{ \bdw_k^H \bdg_i }^2 - 2 \Re \left\{ \bdw_k^H \bdg_k \cdot b_k^{(n)} \right\} \right) \\
	\mathrm{s.t.}\  & \sum_{k=1}^{K} \norm{ \bdw_k }^2 \le P. 
\end{align}
\end{subequations}
The optimal solution to $\clS^{(n)}$ can be easily derived by minimizing its Lagrangian function. Thus, the precoders $\bdW^{(n+1)}$ are given by
\begin{equation}
\bdw_k^{(n+1)} = \left( \sum_{i=1}^{K} a_i^{(n)} \bdg_i \bdg_i^H + \mu^{(n)} \bdI_M \right)^{-1} \bdg_k \cdot b_k^{(n)}\comma\  k \in \clK\comma \label{wk_iteration_ergodic}
\end{equation}
where $\mu^{(n)} \ge 0$ is chosen to make $\sum_{k=1}^{K} \smallnorm{\bdw_k^{(n+1)}}^2 = P$. The precoder design algorithm for solving $\clS$ is summarized in \Cref{algorithm_DL_Tx_Design_Inst_Rx}. By taking advantage of the LEO satellite channel peculiarities, we only need to compute the scalar parameters $\{a_k^{(n)},b_k^{(n)}\}_{k=1}^K$ in each iteration.

\begin{algorithm}[!t]
	\caption{Precoder design algorithm for solving $\clS$.} 
	\label{algorithm_DL_Tx_Design_Inst_Rx}
	\begin{algorithmic}[1]
		\REQUIRE Initialize precoding vector $\bdw_k^{(0)} = \bdw_k^{\init}$, $k \in \clK$, iteration index $n = 0$, and maximum number of iterations $\Niter$.
		\ENSURE Precoding vectors $\{\bdw_k\}_{k=1}^K$.
		\WHILE 1
		\STATE Calculate $a_k^{(n)}$ and $b_k^{(n)}$ for all $k \in \clK$.
		\STATE Update $\{\bdw_k^{(n+1)}\}_{k=1}^K$ with \eqref{wk_iteration_ergodic}.
		\IF{$n\ge\Niter-1$ \textbf{or} $\smallabs{ \sum\nolimits_{k=1}^{K} \xiter{\clR_k}{n+1} - \sum\nolimits_{k=1}^{K} \xiter{\clR_k}{n}  } < \epsilon$}
		\STATE Set $\bdw_k:=\bdw_k^{(n+1)}$, $\forall k \in \clK$, \textbf{break}.
		\ELSE
		\STATE Set $n:=n+1$.
		\ENDIF
		\ENDWHILE	
	\end{algorithmic}	
\end{algorithm}

Due to the expectation in the ergodic rate $\clR_k$, the Monte-Carlo method with exhaustive sample average is required to compute the precoding vectors, which is a computational demanding task when a large number of samples are considered on the averaging procedure. In the next section, we will present low-complexity transmit designs that avoid the sample average.

\section{Transmit Designs with Ergodic Sum Rate Upper Bound} \label{Section_Low_Complexity_DL_Precoder_Design}
In this section, to avoid the exhaustive sample average, we propose transmit designs by approximating the ergodic sum rate with its upper bound. We first prove that in this case, the optimal transmit covariance matrices are still rank-one. Therefore, the design of the transmit covariance matrices can also be boiled down to that of the precoding vectors. 
Then, we show that the design of the precoding vectors can be further converted into that of the scalar variables, and we devise an algorithm to compute these scalar variables. For the ease of real-time processing, we further propose a low-complexity solution to calculate the scalar variables based on a learning framework. The proposed learning-based solution can achieve near-optimal performance, which will be demonstrated in the next section.

\subsection{Rank-One Property of Transmit Covariance Matrices} \label{subsec_simplify_precoder_design}
Notice that $f(x) = \log \left( 1 + \frac{ a x }{ b x + c } \right)$ is a concave function of $x \ge 0$ for $a,b,c \ge 0$ \cite{CSun2015BDMA}. By invoking the Jensen's inequality \cite{BoydConvexOptimization}, the DL ergodic rate $\clI_k$ of UT $k$ can be upper bounded by
\begin{align}
\clI_k 
&= \mathbb{E} \left\{ \log \left( 1 + \frac{  \bdg_k^H \bdQ_k \bdg_k \norm{\bdd_k}^2 }{ \sum_{i \ne k} \bdg_k^H \bdQ_i \bdg_k \norm{\bdd_k}^2 + \sigma_k^2 } \right) \right\} \notag \\
&\le \log \left( 1 + \frac{ \bdg_k^H \bdQ_k \bdg_k \beta_k  }{ \sum_{i \ne k} \bdg_k^H \bdQ_i \bdg_k \beta_k  + \sigma_k^2 } \right) \triangleq \clI_k^{\ub}. \label{Rate_UT_k_DL_noRx_UB}
\end{align}
The problem of maximizing the upper bound of the DL ergodic sum rate can be formulated as
\begin{subequations}\label{Problem_SumRate_Max_Covariance_UB}
\begin{align}
\clP^{\ub}: \ \max_{ \left\{ \bdQ_k \right\}_{k=1}^K }\ & \sum_{k=1}^{K} \clI_k^{\ub}\\ 
\mathrm{s.t.}\ & \sum_{k=1}^{K} \trace(\bdQ_k) \le P, \ \bdQ_k \succeq \bdzro, \ \forall k \in \clK.	
\end{align}
\end{subequations}

\begin{mytheorem} \label{Prop_rank-one_Covariance_UB}
	The optimal $\{\bdQ_k\}_{k=1}^K$ to the problem $\clP^{\ub}$ must satisfy $\rank(\bdQ_k) \le 1$, $\forall k \in \clK$. 
\end{mytheorem} 
\begin{IEEEproof}
	The proof is similar with that in \Cref{Prop_rank-one_Covariance}. Thus, it is omitted here.
\end{IEEEproof}

According to \Cref{Prop_rank-one_Covariance_UB}, the rank of the optimal transmit covariance matrices to the problem $\clP^{\ub}$ should be no greater than one, which manifests that the single-stream precoding strategy for each UT suffices to maximize the upper bound on the ergodic sum rate. Thus, we can rewrite the transmit covariance matrix $\bdQ_k$ as $\bdQ_k = \bdw_k \bdw_k^H$, and once more the design of the transmit covariance matrices $\{\bdQ_k\}_{k=1}^K$ can be reduced to that of the precoding vectors $\{\bdw_k\}_{k=1}^K$. Hence, the $\clI_k^{\ub}$ expression in \eqref{Rate_UT_k_DL_noRx_UB} can be further written as
\begin{equation}
\clI_k^{\ub} = \log \left( 1 + \frac{ \abs{\bdw_k^H \bdg_k}^2 \beta_k  }{ \sum_{i \ne k} \abs{\bdw_i^H \bdg_k}^2 \beta_k  + \sigma_k^2 } \right) \triangleq \clR_k^{\ub}. \label{clRk_ub}
\end{equation}
Here, $\clI_k^{\ub}$ is replaced with $\clR_k^{\ub}$, because $\clR_k^{\ub}$ has become a closed-form expression of the precoding vectors $\{\bdw_k\}_{k=1}^K$. Then, the transmit covariance matrix optimization problem $\clP^{\ub}$ in \eqref{Problem_SumRate_Max_Covariance_UB} can be reformulated as
\begin{equation}
\clS^{\ub}: \ \max_{  \bdW  } \ \sum_{k=1}^{K} \clR_k^{\ub}\comma \quad \mathrm{s.t.} \ \sum_{k=1}^{K} \lVert \mathbf{w}_k \rVert^2 \le P.	\label{Problem_DL_Tx_design_Inst_Rx_UB}
\end{equation}
Note that the problem $\clS^{\ub}$ is analogous to the sum rate maximization problem in DL multi-user multiple-input single-output (MU-MISO) channels \cite{Emil2014OptimalBeamforming}.
The optimal precoding vectors to the problem $\clS^{\ub}$ must satisfy $\sum_{k=1}^{K} \norm{\bdw_k}^2 = P$, because any precoding vectors with $\sum_{k=1}^{K} \norm{\bdw_k}^2 < P$ can be scaled up to increase the objective value.

It is worth noting that for the problem $\clS^{\ub}$, the channel parameters $\{\beta_k/\sigma_k^2, \tdbdtheta_k\}_{k=1}^K$ are required at the satellite to compute the precoding vectors, which depend on the location information and average channel power of UTs. When the UPA placement is fixed, the space angle pairs $\{\tdbdtheta_k\}_{k=1}^K$ can be derived from the location information of the satellite and UTs, which can be acquired through the global positioning system (GPS). The satellite can obtain the estimation of $\{\beta_k\}_{k=1}^K$ by exploiting the UL sounding signals and the reciprocity of sCSI \cite{CSun2015BDMA}.

\subsection{Precoding Vector Design} \label{subsec_structure_DL_precoder}
In this subsection, we show that the design of high-dimensional precoding vectors in the problem $\clS^{\ub}$ can be transformed into that of $K$ scalar variables. For the ease of statement, we first formulate an optimization problem as follows
\begin{equation} 
	\clM^{\ub}: \ \max_{\bdlambda} \  \sum_{k=1}^{K} r_k\comma \quad 	
	\mathrm{s.t.}\ \sum_{k=1}^K \lambda_k = P\comma \ \lambda_k\ge0\comma \ \forall k \in \clK\comma
\end{equation}
where $\bdlambda = [\lambda_1 \ \cdots \ \lambda_K]^T \in \Real{K}{1}$ and $r_k$ is a function of $\{\lambda_k\}_{k=1}^K$ given by
\begin{align} 
	r_k(\lambda_1,\dots,\lambda_K) &= \log \det \left( \sum_{i=1}^{K} \frac{\lambda_i \beta_i}{\sigma_i^2} \bdg_i \bdg_i^H + \bdI_M \right) \notag \\
	&\quad - \log \det \left( \sum_{i \ne k} \frac{\lambda_i \beta_i}{\sigma_i^2} \bdg_i \bdg_i^H + \bdI_M \right). \label{rk_expression}
\end{align}
The relationship between the problems $\clS^{\ub}$ and $\clM^{\ub}$ will be established in the following.

Denote $\{\bdw_k^{\opt}\}_{k=1}^K$ and $\{\lambda_k^{\opt}\}_{k=1}^K$ as the optimal solutions to problems $\clS^{\ub}$ and $\clM^{\ub}$, respectively. As described in the following theorem, as long as the scalar variables $\{\lambda_k^{\opt}\}_{k=1}^K$ are known, the precoding vectors $\{\bdw_k^{\opt}\}_{k=1}^K$ can be derived in closed form immediately.
\begin{mytheorem} \label{Prop_Solution_UB_DL_instRx}
	The precoding vectors $\{\bdw_k^{\opt}\}_{k=1}^K$ can be written as
	\begin{equation} 
		\bdw_k^{\opt} = \sqrt{q_k^{\opt}} \cdot \frac{ \Xinv{\bdV^{\opt}} \bdg_k }{ \smallnorm{ \Xinv{\bdV^{\opt}} \bdg_k }  }\comma \ \forall k\in \clK.
		 \label{Optimal_solution_DL_instRx_Jesen}
	\end{equation} 
	In \eqref{Optimal_solution_DL_instRx_Jesen}, the matrix $\bdV^{\opt} \in \Complex{M}{M}$ and $q_k^\opt$ are given by
	\begin{subequations}\label{Solution_wk_V_qk} 
		\begin{align}
			\bdV^\opt &= \sum_{k=1}^K \frac{ \lambda_k^\opt \beta_k }{ \sigma_k^2 } \bdg_k \bdg_k^H + \bdI_M\comma  \label{Solution_wk_V} \\
			q_k^\opt &= \frac{\lambda_k^\opt \beta_k (\gamma_k^\opt+1) }{\mu^{\opt} \sigma_k^2　}  \smallnorm{\Xinv{\bdV^\opt} \bdg_k}^2\comma \label{Solution_wk_qk} 
		\end{align} 
	\end{subequations} 
	where the parameters $\gamma_k^{\opt}$ and $\mu^{\opt}$ in \eqref{Solution_wk_qk} are also determined by $\{\lambda_k^{\opt}\}_{k=1}^K$ as follows
	\begin{subequations} \label{Solution_wk_mu_gamma}
		\begin{align}
			\gamma_k^{\opt} &= \xinv{ 1 - (\lambda_k^{\opt} \beta_k / \sigma_k^2) \bdg_k^H \Xinv{\bdV^{\opt}} \bdg_k } - 1\comma \label{Solution_wk_gamma} \\
			\mu^{\opt} &= \xinv{P} \sum_{k=1}^{K} \frac{\lambda_k^\opt \beta_k (\gamma_k^\opt+1) }{ \sigma_k^2　} \smallnorm{\Xinv{\bdV^\opt} \bdg_k}^2. \label{Solution_wk_mu} 
		\end{align}
	\end{subequations}
\end{mytheorem}
\begin{IEEEproof}
	Please refer to \Cref{appendix_solution_UB_DL_instRx_proof}.
\end{IEEEproof}
 
In massive MIMO LEO SATCOM systems, the dimension of the precoding vectors $\{\bdw_k\}_{k=1}^K$ might be extremely large. 
\Cref{Prop_Solution_UB_DL_instRx} indicates that the design of the high-dimensional precoding vectors $\{\bdw_k\}_{k=1}^K$ in the problem $\clS^{\ub}$ can be simplified into that of $K$ scalar variables $\{\lambda_k\}_{k=1}^K$ in the problem $\clM^{\ub}$, with which the precoding vectors $\{\bdw_k\}_{k=1}^K$ can be calculated in closed-form. 

Next, by resorting to the MM framework, we present an algorithm to compute the scalar variables $\{\lambda_k\}_{k=1}^K$.
First, we replace the non-convex function $r_k$ with one of its concave minoring functions. Then, a locally optimal solution to the problem $\clM^{\ub}$ can be obtained by solving a series of convex programs sequentially.
For given scalar variables $\{\lambda_k^{(n)}\}_{k=1}^K$ in the $n$th iteration, a minoring function of $r_k$ is constructed as follows
\begin{align}
	\xiter{h_k}{n} &= -  \left( \sum_{i=1}^{K} \xiter{\psi_{k,i}}{n} \frac{ \lambda_i \beta_i }{ \sigma_i^2 } - 2 \xiter{\chi_k}{n} \sqrt{ \frac{ \lambda_k \beta_k }{ \sigma_k^2 } }  \right) \notag \\
	&\qquad - \xiter{\delta_k}{n} + 1 + \xiter{r_k}{n}\comma
\end{align}
where $\xiter{\psi_{k,i}}{n}$, $\xiter{\chi_k}{n}$ and $\xiter{\delta_k}{n}$  are shown in \Cref{appendix_minorizing_rk_DL_instRx_proof}. Then, a locally optimal solution to $\clM^{\ub}$ can be obtained by iteratively solving the following convex subproblem
\begin{equation}
	\begin{aligned}
		\clM_n^{\ub}: \ \max_{ \{\lambda_k \ge 0\}_{k=1}^K } \ \sum_{k=1}^{K} \xiter{h_k}{n}\comma \quad \mathrm{s.t.}  \ \sum_{k=1}^K \lambda_k = P\comma
	\end{aligned}
\end{equation}
which is equivalent to
\begin{subequations} \label{Mub_convex_subprobelm}
\begin{align}
	\clM_n^{\ub}: \ \min_{ \{\lambda_k \ge 0\}_{k=1}^K } \ & \sum_{k=1}^{K}  \sum_{i=1}^{K} \xiter{\psi_{i,k}}{n} \frac{ \lambda_k \beta_k }{ \sigma_k^2 } - 2 \sum_{k=1}^{K} \xiter{\chi_k}{n} \sqrt{ \frac{ \lambda_k \beta_k }{ \sigma_k^2 } } \\
	\mathrm{s.t.}\ & \sum_{k=1}^K \lambda_k = P.
\end{align}
\end{subequations}
By applying the Lagrangian minimization method, the optimal solution to $\clM_n^{\ub}$ is given by
\begin{equation}
	\lambda_k^{(n+1)} = \frac{ \left( \chi_k^{(n)} \right)^2 \frac{ \beta_k  }{ \sigma_k^2 } }{ \left( \sum_{i=1}^{K} \psi_{i,k}^{(n)} \frac{\beta_k}{\sigma_k^2} + \nu^{(n)} \right)^2  }\comma \ \forall k \in \clK\comma \label{lambda_Update_DL_Tx_Design_Jesen_Dual}
\end{equation}
where $\nu^{(n)}$ can be obtained by the bisection search method such that  $\sum_{k=1}^{K} \xiter{\lambda_k}{n+1} = P$. The variable $\nu^{(n)}$ must satisfy $\nu^{(n)} \ge - \min\limits_{k \in \clK} \sum_{i=1}^{K} \psi_{i,k}^{(n)} \frac{\beta_k}{\sigma_k^2}$.
After the scalar variables $\{\lambda_k\}_{k=1}^K$ are known, the precoding vectors $\{\bdw_k\}_{k=1}^K$ can be derived with \Cref{Prop_Solution_UB_DL_instRx}. The detailed procedures for solving $\clS^{\ub}$ are summarized in \Cref{algorithm_DL_Tx_Design_Inst_Rx_Jesen_Dual}. 

Notice that the computational complexity in terms of the number of multiplication operations in each iteration of \Cref{algorithm_DL_Tx_Design_Inst_Rx_Jesen_Dual} is given by $K^3 + 2K^2 M$. After the parameters $\{\lambda_k\}_{k=1}^K$ are determined, we need to compute the precoding vectors $\{\bdw_k\}_{k=1}^K$ by using \Cref{Optimal_solution_DL_instRx_Jesen,Solution_wk_V_qk,Solution_wk_mu_gamma}, whose complexity is given by $K^3 + K^2 M$. Thus, the total computational complexity of \Cref{algorithm_DL_Tx_Design_Inst_Rx_Jesen_Dual} can be expressed as $\Niter (K^3 + 2 K^2 M) + K^3 + K^2 M$.

\begin{algorithm}[!t]
	\caption{Precoder design algorithm for solving $\clS^{\ub}$.} 
	\label{algorithm_DL_Tx_Design_Inst_Rx_Jesen_Dual}
	\begin{algorithmic}[1]
		\REQUIRE Initialize scalar variables $\lambda_k^{(0)} = \lambda_k^{\init}(\forall k \in \clK)$, iteration index $n = 0$, and maximum number of iterations $\Niter$.
		\ENSURE Precoding vectors $\{\bdw_k\}_{k=1}^K$.
		\WHILE 1
		\STATE Calculate $\sum_{i=1}^{K} \psi_{i,k}^{(n)}$ and $\chi_k^{(n)}$ for all $k \in \clK$.
		\STATE Update $\{\lambda_k^{(n+1)}\}_{k=1}^K$ with \eqref{lambda_Update_DL_Tx_Design_Jesen_Dual}.
		\IF{$n\ge\Niter-1$ \textbf{or} $\smallabs{ \sum\nolimits_{k=1}^{K} \xiter{r_k}{n+1} - \sum\nolimits_{k=1}^{K} \xiter{r_k}{n}  } < \epsilon$}
		\STATE Set $\lambda_k:= \lambda_k^{(n+1)}$, $\forall k \in \clK$, \textbf{break}.
		\ELSE 
		\STATE Set $n:=n+1$.
		\ENDIF
		\ENDWHILE
		\STATE Compute the precoding vectors $\{\bdw_k\}_{k=1}^K$ with \Cref{Optimal_solution_DL_instRx_Jesen,Solution_wk_V_qk,Solution_wk_mu_gamma}.
	\end{algorithmic}
\end{algorithm}

Although \Cref{algorithm_DL_Tx_Design_Inst_Rx_Jesen_Dual} can be used to compute the precoding vectors to the problem $\clS^{\ub}$, it involves a number of complicated iterations, which renders it challenging to be implemented for real-time signal processing at the limited satellite payloads. 
In the next subsection, we propose a solution based on the learning framework to compute the scalar variables $\{\lambda_k\}_{k=1}^K$. The input and output in the constructed neural network (NN) both have low-dimensional structures, so that the onboard  implementation complexity is reduced significantly.

\subsection{Learning to Compute Scalar Variables $\{\lambda_k\}_{k=1}^K$} \label{subsec_precoder_design_learning}
In the past years, machine learning \cite{Goodfellow2016DeepLearning} has been intensively studied to address the intractable problems in wireless communications, such as the channel estimation \cite{He2018DeepLearningEstimation}, resource allocation \cite{Wang2021NOMACompletionTime}, DL precoder design \cite{Shi2020RobustPrecodingLearning}, etc. 
By using simple linear operations, e.g., matrix-vector multiplications, and nonlinear activation functions as building blocks, machine learning provides a low-complexity way to fit the output of conventional iterative algorithms. 

In this subsection, we elaborate the computation of the scalar variables $\{\lambda_k\}_{k=1}^K$ in the problem $\clM^{\ub}$ with the learning-based approach.
From \Cref{subsec_structure_DL_precoder}, we can see that the optimal scalar variables $\{\lambda_k^{\opt}\}_{k=1}^K$ can be fully determined by the transmit power $P$ and the channel parameters $\{\tdbeta_k,\tdtheta_k^{\rx},\tdtheta_k^{\ry}\}_{k=1}^K$ with $\tdbeta_k = \beta_k/\sigma_k^2$. 
Moreover, we normalize the optimal scalar variables $\{\lambda_k^{\opt}\}_{k=1}^K$ as $\{\tdlambda_k^{\opt}\}_{k=1}^K$ with $\tdlambda_k^{\opt} = \lambda_k^{\opt} K /P$, such that $\tdlambda_k^{\opt}$'s can be in the same order of magnitude for different transmit power. 
It can be assumed that there exists a nonlinear mapping $\clF(\cdot): \Real{(3K+1)}{1} \rightarrow \Real{K}{1}$, which maps $\bdm \triangleq [\tdbeta_1 \ \tdtheta_1^{\rx} \ \tdtheta_1^{\ry}\ \cdots \ \tdbeta_K\ \tdtheta_K^{\rx}\ \tdtheta_K^{\ry}\ P]^T \in \Real{(3K+1)}{1}$ to $\bdv \triangleq [\tdlambda_1^{\opt} \ \cdots \ \tdlambda_K^{\opt}]^T \in \Real{K}{1}$, i.e.,
\begin{equation} 
	\bdv = \clF(\bdm).
\end{equation}

\begin{figure}[h]
	\centering
	\vspace{-0em}
	\includegraphics[width=0.45\textwidth]{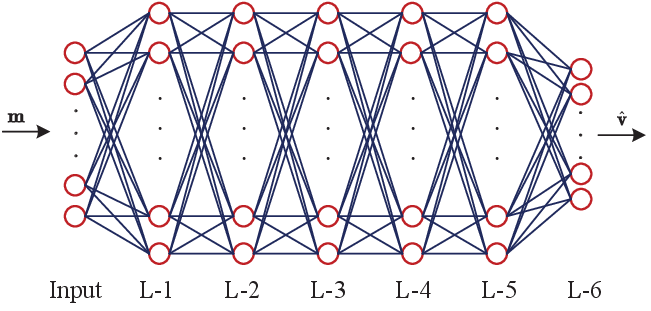}
	\caption{Example of a six-layer perceptron.}
	\label{fig_example_MLP}
	\vspace{-0em}
\end{figure}

NN can approximate the nonlinear mapping $\clF(\cdot)$ with another one $\hat{\clF}(\cdot,\bdomega):\Real{(3K+1)}{1} \rightarrow \Real{K}{1}$ parameterized by $\bdomega$, which maps $\bdm$ to a prediction of $\bdv$, i.e.,
\begin{equation} 
	\hat{\bdv} = \hat{\clF}(\bdm,\bdomega).
\end{equation}
It is expected that the prediction $\hat{\bdv}$ can be as close to the accurate $\bdv$ as possible. 
In this paper, we use the multilayer perceptron (MLP), which is a special class of NN, to learn the nonlinear mapping $\clF(\cdot)$. As shown in \Cref{fig_example_MLP}, the MLP is composed of a number of layers, while each layer in the MLP has plenty of neurons.
We use the rectified linear units (ReLU), i.e., $\mathsf{G}(x) = \max\{x,0\}$, as the activation function to make the output of each layer non-negative. 

Let $\{(\bdm_{i},\bdv_{i})\}_{i=1}^{N_D}$ denote the dataset, where $\bdm_i$ and $\bdv_i$ represent the $i$th samples of $\bdm$ and $\bdv$, respectively, and $N_D$ is the number of samples in the dataset.
We choose the MSE as the loss function for training the NN, i.e.,
\begin{equation} 
	\text{Loss} = \xinv{N_D} \sum_{i=1}^{N_D} \smallnorm{ \bdv_{i} - \hat{\bdv}_{i} }^2\comma
\end{equation}
where $\hat{\bdv}_i$ is the prediction of $\bdv_i$. 
The training stage can be performed offline at the ground station. When the training stage is complete, the ground station needs to feed the parameters of the NN back to the satellite.
In the testing stage, to make the output of the NN a feasible solution to the problem $\clM^{\ub}$, we normalize the prediction $\hat{\bdv}$ as follows
\begin{equation} 
	\htbdlambda = \frac{P}{\bdone^T \hat{\bdv}} \hat{\bdv}\comma
\end{equation} 
such that $\htbdlambda \in \Real{K}{1}$ satisfies $\bdone^T \htbdlambda = P$. Then, $\htbdlambda$ is treated as the prediction of the optimal solution to the problem $\clM^{\ub}$ through the NN. With the obtained scalar variables in $\htbdlambda$, the precoding vectors can be calculated by using \Cref{Optimal_solution_DL_instRx_Jesen,Solution_wk_V_qk,Solution_wk_mu_gamma}. To make the NN adapt to the cases that the number of antennas $M$ or the number of UTs $K$ is changed, the dataset augmentation and transfer learning techniques can be used \cite{Xia2020ModelBeamformingNN}.

The NN involves only rather simple operations, e.g., matrix-vector multiplication followed by an activation function, which have much lower implementation complexity compared with \Cref{algorithm_DL_Tx_Design_Inst_Rx_Jesen_Dual}.
The complexity of the NN regarding to the number of multiplications is $(3K+1) D_1 + \sum_{i=1}^{N_L-1} D_i D_{i+1}$ operations, where $N_L$ is the number of layers and $D_i$ is the number of neurons in the $i$th layer. Since we need to compute the precoding vectors after the scalar variables in $\htbdlambda$ are obtained, the total complexity of the NN-based approach is evaluated by $(3K+1) D_1 + \sum_{i=1}^{N_L-1} D_i D_{i+1} + K^3 + K^2 M$.
It is worth noting that the dimensions of the input and output of the NN are only $3K+1$ and $K$, respectively, which are independent of the number of antennas at the satellite and UTs. Thus, the NN presents higher gains in computational complexity for relatively large $M$ and $K$, which makes it an attractive solution for massive MIMO LEO SATCOM systems. 

\section{Simulation Results} \label{Sectioin_Simulation}

\begin{table}[!t] 
\centering
\footnotesize
\vspace{-0em}
\captionof{table}{Simulation Parameters}
\label{table_simulation}
\begin{tabular}{Lc}
	\toprule
	Parameters & Values \\
	\midrule
	Earth radius $R_e$ & $6378$ km \\
	Orbit altitude $H$ & $1000$ km \\
	Central frequency $f_c$  & $4$ GHz \\
	Bandwidth $B$ & $50$ MHz	\\
	Noise temperature $T_\rn$ & $290$ K \\
	Number of antennas $\Mx$, $\My$, $\Nx$, $\Ny$ & $12$, $12$, $6$, $6$ \\
	Antenna spacing $d_{\rx}$, $d_{\ry}$, $d_{\rx'}$, $d_{\ry'}$  & $\lambda$, $\lambda$, $\frac{\lambda}{2}$, $\frac{\lambda}{2}$ \\
	Per-antenna gain $G_{\sat}$, $G_{\ut}$ & $6$ dBi, $0$ dBi \\
	Maximum nadir angle $\vtheta_{\max}$ & $\xdeg{30}$ \\
	Number of UTs $K$ & $100$ \\
	Transmit power $P$ & $10$ dBW -- $25$ dBW \\
	Number of layers in NN	& 9 \\
	Number of neurons $D_1$--$D_9$ & $512$ ($D_1$--$D_8$), $100$ ($D_9$) \\
	Dataset size $N_D$ & $2\times 10^5$ ($10\%$ for testing) \\
	Batch size & 128 \\
	Optimizer & Adam \\
	Learning rate & 0.001 \\
	\bottomrule
\end{tabular}
\vspace{-0em}
\end{table}

\begin{figure}[!t]
	\centering
	\vspace{-0em}
	\includegraphics[width=0.45\textwidth]{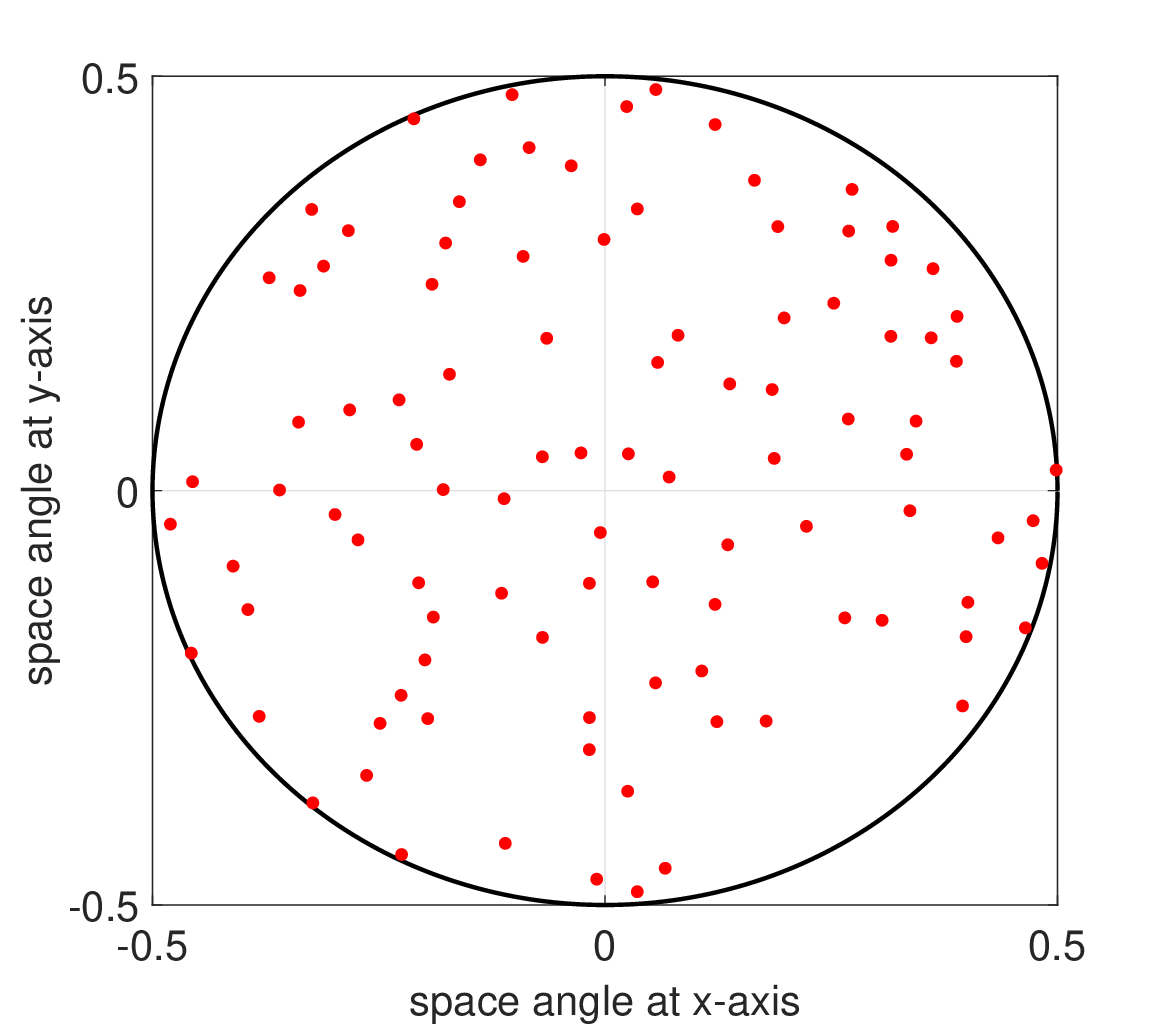} 
	\caption{Distribution of space angle pairs for all UTs.}
	\label{Fig_User_space_angle}
	\vspace{-0em}
\end{figure}

In this section, we present the simulation results to verify the performance of the proposed DL transmit designs in a massive MIMO LEO SATCOM system. 
The simulation parameters are summarized in \Cref{table_simulation}.
The maximum nadir angle of the UTs is denoted as $\vtheta_{\max}$. The space angle pair $\tdbdtheta_k = (\tdtheta_k^{\rx},\tdtheta_k^{\ry})$ should satisfy $(\tdtheta_k^{\rx})^2 + (\tdtheta_k^{\ry})^2 \le \sin^2 \vtheta_{\max}$ due to the relation $\cos \vtheta_k = \sin \theta_k^{\ry} \sin \theta_k^{\rx} = \sqrt{1 - (\tdtheta_k^{\ry})^2 - (\tdtheta_k^{\rx})^2} \ge \cos \vtheta_{\max}$. 
In the simulations, the Poisson disk sampling \cite{Robert2007PoissonDisk} is used to generate the space angle pairs of UTs within the circle region $\{(x,y): x^2 + y^2 \le \sin^2 \vtheta_{\max}\}$ as shown in \Cref{Fig_User_space_angle} with a minimum distance between any two pairs of space angles given by $\rho_{\min} = 0.037$, which guarantees at least $3$ dB interference power decay among UTs. 
The per-antenna gains at the satellite and UTs are denoted as $G_{\sat}$ and $G_{\ut}$, respectively.
For simplicity, we assume that each antenna element at the satellite has the ideal directional power pattern $R(\theta_{\rx},\theta_{\ry}) = G_{\sat}$, if $ (\sin\theta_{\ry} \cos \theta_{\rx})^2 + (\cos \theta_{\ry})^2 \le \sin^2 \vtheta_{\max}$, and otherwise, $R(\theta_{\rx},\theta_{\ry}) = 0$, which is in accord with the coverage area seen at the satellite. 
The elevation angle of UT $k$ in \Cref{fig_UPA} can be computed by $\alpha_k = \cos^{-1} \left( \frac{R_s}{R_e} \sin \vtheta_k \right)$ \cite{Lutz2000SatSysPerson}, where $R_e$ is the earth radius, $R_s = R_e + H$ is the orbit radius.
The distance between the satellite and UT $k$ in \Cref{fig_UPA} is given by $D_k = \sqrt{R_e^2 \sin^2 \alpha_k + H^2 + 2 H R_e} - R_e \sin \alpha_k$ \cite{3GPP_NonTerrestrial}.
The random vector $\bdd_k = \sqrt{\frac{\kappa_k \beta_k}{\kappa_k + 1}} \bdd_{k,0} + \sqrt{\frac{\beta_k}{\kappa_k + 1}} \tdbdd_k$ in \eqref{channel_matrix_UTk} is simulated in terms of $\bdd_k(t,f)$ in \eqref{channel_model_UTk_compensate}, where the first path is used to produce the LoS direction $\bdd_{k,0} = \bdd(\bdvphi_{k,0})$ and the remaining $L_k-1$ paths are used for $\tdbdd_k$. For simplicity, each UT's UPA is assumed to be placed horizontally, which implies that $\bdvphi_{k,0}$ satisfies $\sin\vphi_{k,0}^{\ry'} \sin \vphi_{k,0}^{\rx'} = \sin \alpha_k$ (e.g., $\vphi_{k,0}^{\rx'} = \xdeg{90}$ and $ \vphi_{k,0}^{\ry'} = \alpha_k$). 
To simulate $\tdbdd_k$, the path gains $\{a_{k,\ell}\}_{\ell=1}^{L_k-1}$ are generated by using the exponential power delay profile, while the paired AoAs $\{\bdvphi_{k,\ell}\}_{\ell=1}^{L_k-1}$ are produced according to the wrapped Gaussian power angle spectrum, as described in the 3GPP technical report on non-terrestrial networks \cite[Section 6]{3GPP_NonTerrestrial}. Moreover, the pathloss, shadow fading and Rician factors are computed in accordance with the suburban scenarios, and the ionospheric loss  is set as $1$ dB approximately \cite[Section 6]{3GPP_NonTerrestrial}.
The average channel power $\beta_k$ is simulated by $\xinv{N_S} \sum_{n=1}^{N_S} \smallnorm{ \bdd_{k,n} }^2$, where $\bdd_{k,n}$ is the $n$th sample of $\bdd_k$ and the number of channel samples is set as $N_S = 1000$.
The noise variance is given by $\sigma_k^2 = k_\rB T_\rn B$ where $k_\rB = 1.38 \times 10^{-23} \text{ J} \cdot \text{K}^{-1}$ is the Boltzmann constant, $T_{\rn}$ is the noise temperature and $B$ is the system bandwidth.

In order to demonstrate the performance of the NN-based approach, we use \Cref{algorithm_DL_Tx_Design_Inst_Rx_Jesen_Dual} to generate the dataset.  The dataset for training and testing the NN is available at GitHub: \underline{\textit{https://github.com/likexin1415}}. Besides, we use the TensorFlow toolbox to train the NN. The structure of the NN, dataset size, batch size, optimizer and learning rate are also presented in \Cref{table_simulation}. 
%The samples with higher transmit power are assigned with larger weights in the loss function. 
%Here, we set the weight $w_i$ as $w_i = \lfloor9^{2\lg P_i - 5} + 1\rfloor$, where $P_i$ (in Watt) is the transmit power for the $i$th sample.

\begin{figure}[!t]
	\centering
	\vspace{-0em}
	\includegraphics[width=0.45\textwidth]{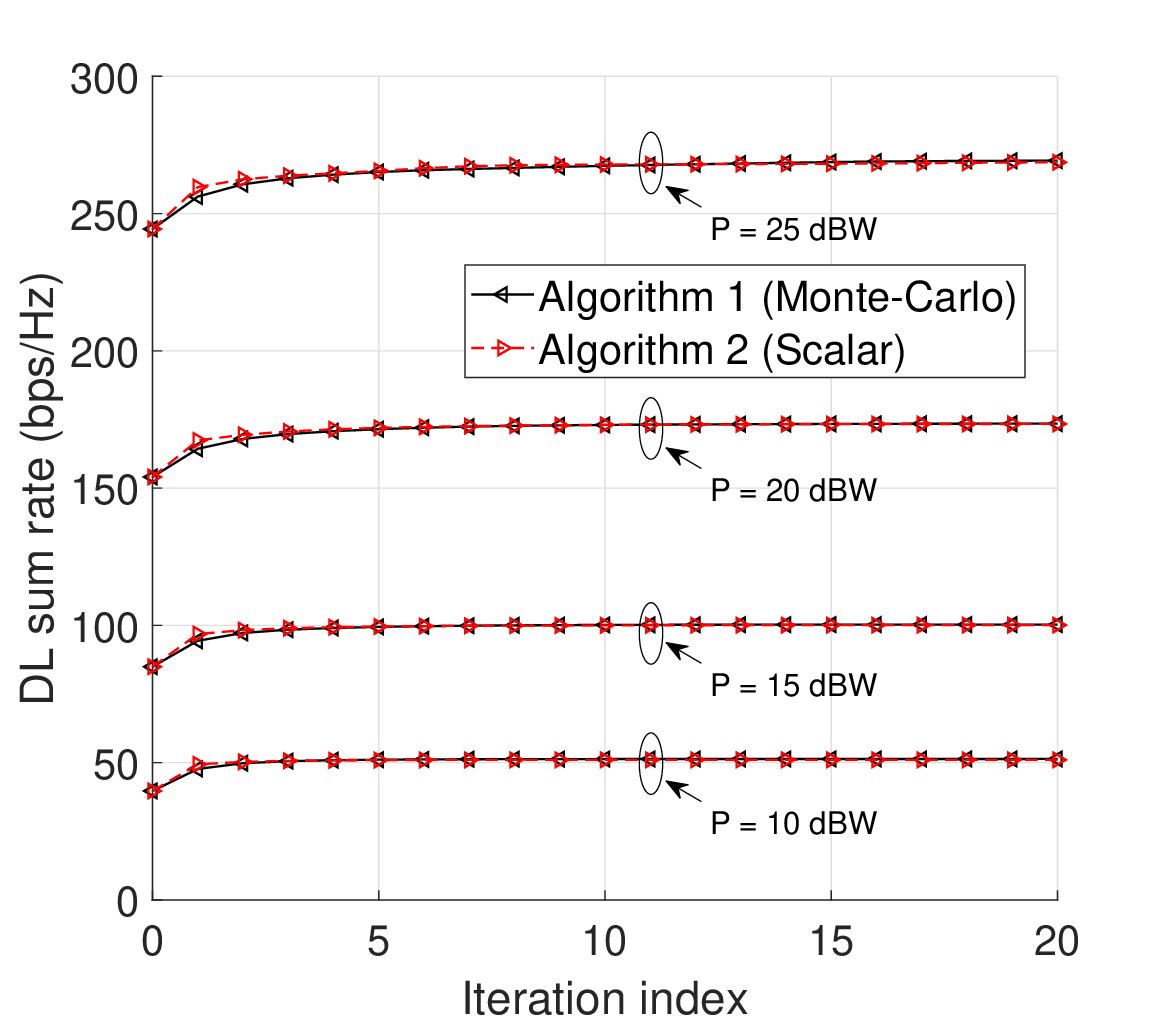}
	\caption{Convergence of \Cref{algorithm_DL_Tx_Design_Inst_Rx,algorithm_DL_Tx_Design_Inst_Rx_Jesen_Dual}.}
	\label{Fig_DL_Convergence}
	\vspace{-0em}
\end{figure}

In \Cref{Fig_DL_Convergence}, the convergence performance of \Cref{algorithm_DL_Tx_Design_Inst_Rx,algorithm_DL_Tx_Design_Inst_Rx_Jesen_Dual} is shown.
It is observed that \Cref{algorithm_DL_Tx_Design_Inst_Rx,algorithm_DL_Tx_Design_Inst_Rx_Jesen_Dual} converge within about $20$ times of iterations.
Hence, in the simulations, the maximum number of iterations $\Niter$ is set as $\Niter = 20$ for both \Cref{algorithm_DL_Tx_Design_Inst_Rx,algorithm_DL_Tx_Design_Inst_Rx_Jesen_Dual}. 
By using the NN parameters as shown in \Cref{table_simulation}, the complexity of the NN-based approach is only about $5.6\%$ of that of  \Cref{algorithm_DL_Tx_Design_Inst_Rx_Jesen_Dual}. 

\begin{figure}[!t]
	\centering
	 \vspace{-0em}
	\includegraphics[width=0.45\textwidth]{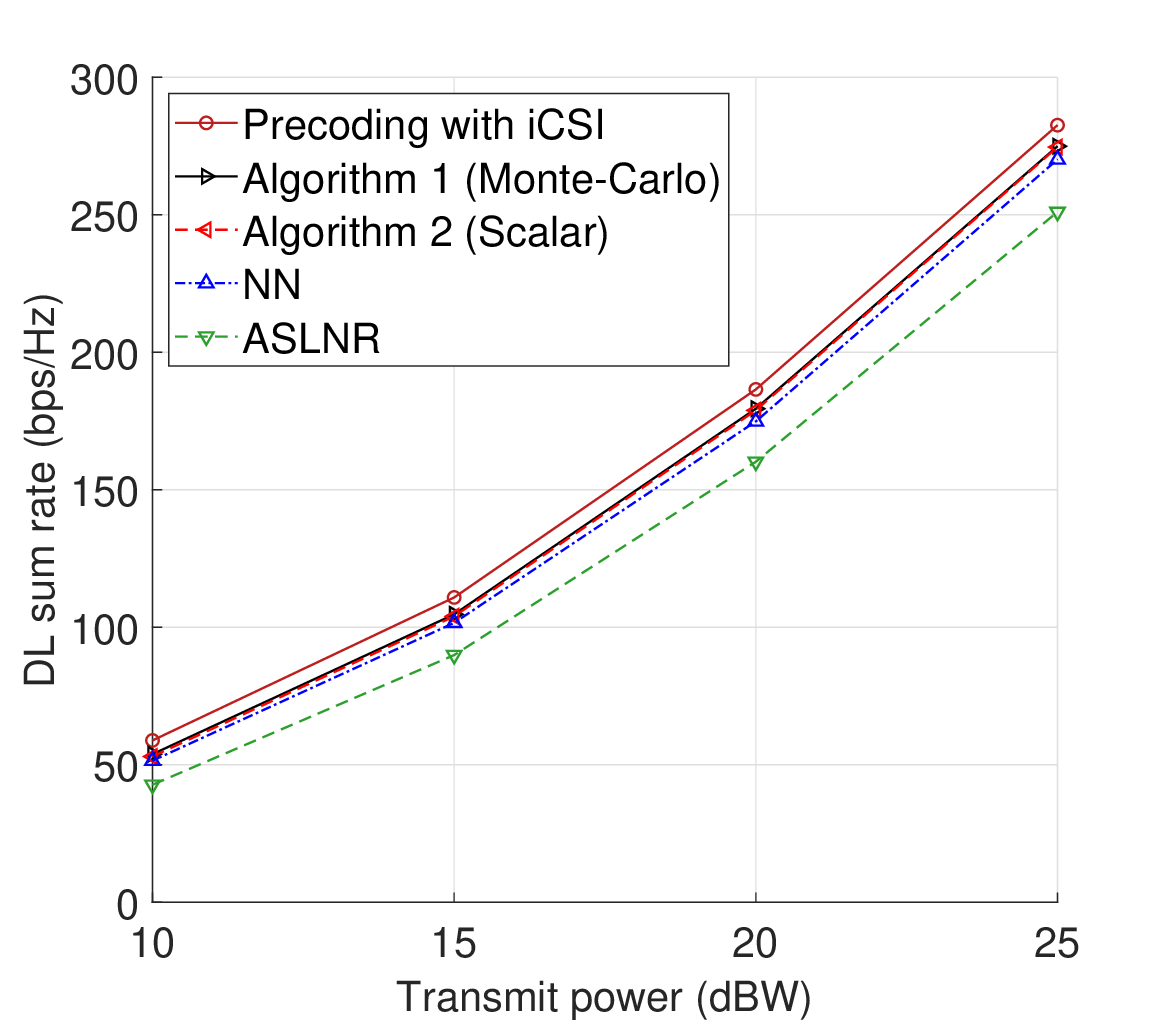}
	\caption{DL sum rate performance of  \Cref{algorithm_DL_Tx_Design_Inst_Rx,algorithm_DL_Tx_Design_Inst_Rx_Jesen_Dual}, and NN.}
	\label{Fig_DL_Performance}
	\vspace{-0em}
\end{figure}

In \Cref{Fig_DL_Performance}, the sum rate performance of \Cref{algorithm_DL_Tx_Design_Inst_Rx,algorithm_DL_Tx_Design_Inst_Rx_Jesen_Dual}, and the NN-based approach is depicted. The performance for the precoding scheme with perfect iCSIT derived from the MM algorithm is also illustrated in \Cref{Fig_DL_Performance}. It is shown that \Cref{algorithm_DL_Tx_Design_Inst_Rx,algorithm_DL_Tx_Design_Inst_Rx_Jesen_Dual}, as well as the NN-based approach, with only sCSIT can achieve close performance to that of the precoding scheme with iCSIT. We can see that the difference of the sum rate performance between \Cref{algorithm_DL_Tx_Design_Inst_Rx,algorithm_DL_Tx_Design_Inst_Rx_Jesen_Dual} is negligible, and the  NN-based approach can achieve near-optimal performance with much lower computational complexity.  In addition, the performance of the ASLNR precoding vectors $\{ \bdw_k^{\aslnr} \}_{k=1}^K$ in the previous work \cite{You2019MassiveMIMOLEO} is also shown for comparison, where $\bdw_k^{\aslnr} = \sqrt{p_k} \cdot \frac{\bdT_k^{-1} \bdg_k}{\smallnorm{\bdT_k^{-1} \bdg_k}} $, $\bdT_k = \sum_{i=1}^{K} \beta_i \bdg_i \bdg_i^H + \frac{\sigma_k^2}{p_k} \bdI_M $, and the power $p_k$ is set as $p_k = \frac{P}{K}$ for simplicity. 
The NN-based approach also shows better performance compared with the ASLNR precoding vectors, which have almost $1$ dB performance loss at $P=25$ dBW. Since the proposed precoding vector design approaches only rely on the slow-varying sCSI, which is independent of subcarriers and OFDM symbols within a stable sCSI period, the onboard implementation complexity could be pretty low. Therefore,  the proposed approaches provide practical solutions for high-throughput massive MIMO LEO SATCOM systems.

\section{Conclusion} \label{Section_Conclusion}

In this paper, we have investigated the DL transmit design with sCSIT in massive MIMO LEO SATCOM systems. First, we derived the DL massive MIMO LEO satellite channel model, where the satellite and the UTs are both equipped with UPAs. Then, we showed that the single-stream precoding for each UT is able to maximize the ergodic sum rate for the linear transmitters. Afterwards, we devised an algorithm to compute the precoding vectors by concentrating on the ergodic sum rate maximization. To reduce the computational complexity, we formulated another transmit design by using an upper bound on the ergodic sum rate, for which the optimality of single-stream precoding also holds. Moreover, we revealed that the design of precoding vectors can be simplified into that of scalar variables, for which an effective algorithm was developed. Furthermore, we proposed a learning-based solution to compute the scalar variables, which involves much lower implementation complexity than iterative algorithms.  Finally, the effectiveness and the performance gains of the proposed DL transmit designs were verified via the simulation results.

\appendices

\section{Proof of \Cref{Prop_rank-one_Covariance}} \label[secinapp]{appendix_rank-one_Covariance_proof}
Our proof is in the spirit of the results in \cite{Sun2019BDMAOptical}.
We first prove that the optimal solution to $\clP$ must be of rank-one. Then, the proof steps can be directly applied to the optimal solution to $\clP^{\ub}$.
The gradient of $\clR_i$ with respect to $\bdQ_k$ can be calculated by
\begin{equation}
\begin{aligned}
\frac{ \partial \clR_i }{ \partial \bdQ_k^T } =
\begin{cases}
	\bbE\left\{ \frac{ \norm{ \bdd_k }^2 }{ T_{k}(\bdd_k) } \right\} \bdg_k \bdg_k^H\comma & \text{ if } i = k\comma \\
	\left( \bbE\left\{ \frac{ \norm{ \bdd_i }^2 }{ T_{i}(\bdd_i) } \right\} - \bbE \left\{ \frac{ \norm{ \bdd_i }^2 }{ I_{i}(\bdd_i) } \right\} \right) \bdg_i \bdg_i^H\comma & \text{ if } i \ne k\comma
\end{cases}
\end{aligned}
\end{equation}
where $T_{k}(\bdd_k) = \sigma_k^2 + \sum_{\ell=1}^K \bdg_k^H \bdQ_{\ell} \bdg_k \norm{\bdd_k}^2$ and $I_{k} (\bdd_k) = \sigma_k^2 + \sum_{\ell \ne k} \bdg_k^H \bdQ_{\ell} \bdg_k \norm{\bdd_k}^2$.
The Lagrangian function of $\clP$ is given by
\begin{equation}
\clL_{\clP} = \sum_{k=1}^{K} \clR_k - v \left( \sum_{k=1}^{K} \trace \left( \bdQ_k \right) - P \right) + \sum_{k=1}^{K} \trace \left( \bdPhi_k \bdQ_k \right)\comma
\end{equation}
where $v \ge 0$ and $\bdPhi_k \succeq \bdzro$ are the Lagrange multipliers associated with the power constraint $\sum_{k=1}^{K} \trace \left( \bdQ_k \right) \le P$ and the positive semidefinite matrix constraint $\bdQ_k \succeq \bdzro$. From the Karush-Kuhn-Tucker (KKT) conditions, the gradient of $\clL_{\clP}$ with respect to the optimal $\bdQ_k$ should be zero, i.e.,
\begin{equation}
\frac{ \partial \clL_{\clP} }{ \partial \bdQ_k^T } = - \bdA_k   + \bdB_k - v \bdI_M + \bdPhi_k = \bdzro\comma \label{Grad_zero_LP}
\end{equation}
where $\bdA_k = \sum_{i \ne k} \left( \bbE\left\{ \frac{ \norm{ \bdd_i }^2 }{ I_{i}(\bdd_i) } \right\} - \bbE \left\{ \frac{ \norm{ \bdd_i }^2 }{ T_{i}(\bdd_i) } \right\} \right) \bdg_i \bdg_i^H$ and $\bdB_k = \bbE\left\{ \frac{ \norm{ \bdd_k }^2 }{ T_{k}(\bdd_k) } \right\} \bdg_k \bdg_k^H$ are both positive semidefinite matrices.
%\begin{subequations}
%	\begin{align}
%	\bdA_k &= \xinv{\log 2} \cdot \sum_{i \ne k} \left( \bbE\left\{ \frac{ \norm{ \bdd_i }^2 }{ I_{i}(\bdd_i) } \right\} - \bbE \left\{ \frac{ \norm{ \bdd_i }^2 }{ T_{i}(\bdd_i) } \right\} \right) \bdg_i \bdg_i^H \\
%	\bdB_k &= \xinv{\log 2} \cdot \bbE\left\{ \frac{ \norm{ \bdd_k }^2 }{ T_{k}(\bdd_k) } \right\} \bdg_k \bdg_k^H\comma
%	\end{align}
%\end{subequations}
%respectively.
From \eqref{Grad_zero_LP}, $\bdPhi_k$ can be expressed as $\bdPhi_k = v \bdI_M + \bdA_k - \bdB_k$.
% \begin{equation}
% 
% \bdPhi_k = v \bdI_M + \bdA_k - \bdB_k.
% \end{equation}
To guarantee $\bdPhi_k \succeq \bdzro$, we must have $v>0$.  Thus, we have $\rank(v\bdI_M + \bdA_k) = M$. From the rank-sum inequality $\abs{ \rank(\bdA) - \rank(\bdB) } \le \rank(\bdA+\bdB)$ \cite[0.4.5(d)]{Horn2013MatrixAnalysis}, the rank of $\bdPhi_k$ must satisfy $\rank(\bdPhi_k) \ge M-1$. Due to the Sylvester inequality $\rank(\bdA) + \rank(\bdB) - n \le \rank(\bdA\bdB)$ \cite[0.4.5(c)]{Horn2013MatrixAnalysis}, where $n$ is the column number of $\bdA$, we can obtain
\begin{equation}
\rank(\bdPhi_k) + \rank(\bdQ_k) - M \le \rank(\bdPhi_k \bdQ_k) \stackeq{a} 0\comma
\end{equation} 
where (a) follows from the complementary slackness condition $\bdPhi_k \bdQ_k = \bdzro$. The rank of $\bdQ_k$ will satisfy $\rank(\bdQ_k) \le 1$. This concludes the proof.

\section{A minorizing function of $\clR_k$ } \label[secinapp]{appendix_minorize_udR_DL_instRx_proof}
By using the recovered data symbol $\hat{s}_k$ in \eqref{hat_sk}, we can derive the mean-square error (MSE) for UT $k$ as 
\begin{align}
\MSE_k &= \mathbb{E} \left\{ \left\lvert \hat{s}_k - s_k \right\rvert^2 \right\} \notag \\
&= \sum_{i=1}^K \left\lvert \bdw_i^H \bdg_k \right\rvert^2 \left\lvert \bdc_k^H \bdd_k \right\rvert^2 + \sigma_k^2 \left\lVert \bdc_k \right\rVert^2 \notag \\
&\qquad - 2 \Re \left\{ \bdg_k^H \bdw_k \cdot \bdc_k^H \bdd_k \right\} + 1. \label{MSE_DL}
\end{align}
The linear receiver $\bdc_k$ that minimizes $\MSE_k$ is given by
\begin{align}
\bdc_k^{\mmse} &= \arg \min_{\bdc_k} \MSE_k \notag \\
&= \frac{ \bdg_k^H \bdw_k }{ \sigma_k^2 + \sum_{i=1}^K \abs{ \bdw_i^H \bdg_k }^2 \norm{\bdd_k}^2 } \cdot \bdd_k. \label{Receiver_MMSE_DL}
\end{align}
The MMSE of UT $k$ achieved by $\bdc_k^{\mmse}$ is given by 
\begin{align}
\MMSE_k &= 1 - \frac{ \abs{ \bdw_k^H \bdg_k }^2 \norm{\bdd_k}^2 }{ \sigma_k^2 + \sum_{i=1}^{K} \abs{ \bdw_i^H \bdg_k }^2 \norm{\bdd_k}^2  } \notag \\
&= \xinv{1+\USINR_k}.
\end{align} 
Thus, $\clR_k$ can be rewritten as $\clR_k = \bbE \left\{ \log \left( 1 + \USINR_k \right) \right\} = - \bbE \left\{ \log \MMSE_k \right\}$.
Given the precoding vectors in the $n$th iteration $\bdW^{(n)} = [ \bdw_1^{(n)} \cdots \bdw_K^{(n)} ]$, the MMSE in the $n$th iteration is given by 
$\xiter{\MMSE_k}{n} = \MMSE_k\rvert_{\bdw_k=\bdw_k^{(n)},\forall k\in\clK}$.
From the concavity of $\log(\cdot)$, we can derive a minoring function of $\clR_k$ as 
\begin{align}
\clR_k &\ge \xiter{\clR_k}{n} - \bbE \left\{ \frac{ \MMSE_k - \xiter{\MMSE_k}{n} }{ \xiter{\MMSE_k}{n} } \right\} \notag \\
&\stackgeq{a} \xiter{\clR_k}{n} + 1 - \bbE \left\{ \frac{ \MSE_k }{ \xiter{\MMSE_k}{n} } \right\} \triangleq \xiter{g_k}{n}\comma	\label{DL_udR_k_concavity}
\end{align}
where $\xiter{\clR_k}{n}$ is the DL ergodic rate of UT $k$ in the $n$th iteration, (a) follows from the inequality $\MMSE_k \le \MSE_k$. Note that $\MSE_k$ is a function of the precoding vectors in $\bdW$ and the linear receiver $\bdc_k$. To make the inequality $\clR_k \ge \xiter{g_k}{n}$ hold with equality at $\bdW^{(n)}$, the receiver $\bdc_k$ in $\MSE_k$ should be given by 
$\bdc_k^{(n)} = \bdc_k^{\mmse}\rvert_{\bdw_k=\bdw_k^{(n)},\forall k\in\clK}$.
After substituting $\bdc_k^{(n)}$ into $\MSE_k$, we have
\begin{align}
 \bbE \left\{ \frac{ \MSE_k }{ \xiter{\MMSE_k}{n} } \right\}
&= a_k^{(n)} \sum_{i=1}^K \smallabs{ \bdw_i^H \bdg_k }^2 \notag \\
&\qquad - 2 \Re \left\{ \bdw_k^H \bdg_k \cdot b_k^{(n)} \right\} + c_k^{(n)}\comma
\end{align}
where $a_k^{(n)} = \bbE \left\{ \frac{ \smallabs{\bdd_k^H \bdc_k^{(n)} }^2 }{ \xiter{\MMSE_k}{n} } \right\}$, $b_k^{(n)} = \bbE \left\{ \frac{ \bdd_k^H \bdc_k^{(n)} }{ \xiter{\MMSE_k}{n} } \right\}$ and $c_k^{(n)}= \bbE \left\{ \frac{ \sigma_k^2 \smallnorm{ \bdc_k^{(n)} }^2 + 1 }{ \xiter{\MMSE_k}{n} } \right\}$.

\section{Proof of \Cref{Prop_Solution_UB_DL_instRx}} \label[secinapp]{appendix_solution_UB_DL_instRx_proof}
The problem $\clS^{\ub}$ can be reformulated as
\begin{subequations}	
	\begin{align}
		\clS_1^{\ub}:\ \max_{\bdw_k,\gamma_k,\forall k}\ & \sum_{k=1}^{K} \log (1 + \gamma_k) \label{Problem_precoding_sum_rk_S1} \\
		\mathrm{s.t.} \ & \gamma_k \le \frac{ \abs{\bdw_k^H \bdg_k}^2 \beta_k  }{ \sum_{i \ne k} \abs{\bdw_i^H \bdg_k}^2 \beta_k  + \sigma_k^2 }\comma \ \forall k \label{Problem_precoding_gammak_S1} \\ 
		& \sum_{k=1}^{K} \lVert \mathbf{w}_k \rVert^2 \le P.	\label{Problem_precoding_power_S1}
	\end{align}
\end{subequations}	
Withe some rearrangement to the constraints in \eqref{Problem_precoding_gammak_S1}, problem $\clS_1^{\ub}$ can be rewritten as
\begin{subequations}	
	\begin{align}
		\clS_2^{\ub}:\ \max_{ \bdw_k,\gamma_k,\forall k}\  & \sum_{k=1}^{K} \log (1 + \gamma_k) \label{Problem_precoding_sum_rk_S2} \\
		\mathrm{s.t.}\  & \frac{\beta_k}{\gamma_k \sigma_k^2} \abs{\bdw_k^H \bdg_k}^2 \ge \sum_{i \ne k} \frac{\beta_k}{\sigma_k^2} \abs{\bdw_i^H \bdg_k}^2+ 1\comma \ \forall k\label{Problem_precoding_gammak_S2} \\ 
		& \sum_{k=1}^{K} \lVert \mathbf{w}_k \rVert^2 \le P.	\label{Problem_precoding_power_S2}
	\end{align}
\end{subequations}
The Lagrangian function for the problem $\clS_2^{\ub}$ is given by
\begin{align}
	\clL_{\clS_2^{\ub}} ={}& \sum_{k=1}^{K} \log (1 + \gamma_k) + \sum_{k=1}^K \cklambda_k \left( \frac{\beta_k}{\gamma_k \sigma_k^2} \abs{\bdw_k^H \bdg_k}^2 \right. \notag \\
	& \left. - \sum_{i \ne k} \frac{\beta_k}{\sigma_k^2} \abs{\bdw_i^H \bdg_k}^2  - 1 \right) - \mu \left( \sum_{k=1}^{K} \lVert \mathbf{w}_k \rVert^2 - P \right)\comma
\end{align}
where $\cklambda_k \ge 0$ and $\mu \ge 0$ are the Lagrange multipliers associated with the constraints in \eqref{Problem_precoding_gammak_S2} and \eqref{Problem_precoding_power_S2}, respectively. 
We denote $\{\bdw_k^{\circ},\gamma_k^{\circ}\}_{k=1}^K$ as the optimal solution to the problem $\clS_2^{\ub}$. At the optimum to the problem $\clS_2^{\ub}$, i.e., $\{\bdw_k,\gamma_k\}_{k=1}^K = \{\bdw_k^{\circ},\gamma_k^{\circ}\}_{k=1}^K$, there must exist some Lagrange multipliers $\{\cklambda_k^{\circ}\}_{k=1}^K$ and $\mu^{\circ}$ such that
\begin{subequations}
	\begin{align}
		\left. \frac{\partial \clL_{\clS_2^{\ub}}}{\partial \bdw_k^*} \right|_{\bdw_k = \bdw_k^{\circ}} &= \frac{\cklambda_k^{\circ}\beta_k}{\gamma_k^{\circ} \sigma_k^2} \bdg_k \bdg_k^H \bdw_k^{\circ} - \sum_{i \ne k} \frac{\cklambda_i^{\circ} \beta_i}{\sigma_i^2} \bdg_i \bdg_i^H \bdw_k^{\circ} \notag \\
		&\qquad \qquad \qquad \qquad \quad - \mu^{\circ} \bdw_k^{\circ} = \bdzro\comma \label{grad_Lagrange_wk_S2} \\
		\left. \frac{\partial \clL_{\clS_2^{\ub}}}{\partial \gamma_k} \right|_{\gamma_k = \gamma_k^{\circ}} &= \xinv{1+\gamma_k^{\circ}} - \frac{\cklambda_k^{\circ} \beta_k}{(\gamma_k^{\circ})^2 \sigma_k^2} \abs{\bdg_k^H \bdw_k^{\circ}}^2 = 0. \label{grad_Lagrange_gammak_S2} 
	\end{align}
\end{subequations}
The condition in \eqref{grad_Lagrange_wk_S2} reveals the direction of the precoding vector $\bdw_k^{\circ}$, while its power is reflected in \eqref{grad_Lagrange_gammak_S2}.
From \eqref{grad_Lagrange_wk_S2}, we can see that
\begin{equation}
	\frac{\cklambda_k^{\circ}\beta_k}{\gamma_k^{\circ} \sigma_k^2} \bdg_k \bdg_k^H \bdw_k^{\circ} = \sum_{i \ne k} \frac{\cklambda_i^{\circ} \beta_i}{\sigma_i^2} \bdg_i \bdg_i^H \bdw_k^{\circ} + \mu^{\circ} \bdw_k^{\circ}. \label{wk_direction_inek_mu}
\end{equation}
We divide \eqref{wk_direction_inek_mu} with $\mu^{\circ}$ and obtain that
\begin{equation}
	\frac{\lambda_k^{\circ}\beta_k}{\gamma_k^{\circ} \sigma_k^2} \bdg_k \bdg_k^H \bdw_k^{\circ} = \sum_{i \ne k} \frac{\lambda_i^{\circ} \beta_i}{\sigma_i^2} \bdg_i \bdg_i^H \bdw_k^{\circ} + \bdw_k^{\circ}\comma \label{wk_direction_inek}
\end{equation}
where $\lambda_k^{\circ} = \cklambda_k^{\circ}/\mu^{\circ}$.
After that, we add the term $(\lambda_k^{\circ} \beta_k / \sigma_k^2) \bdg_k \bdg_k^H \bdw_k^{\circ}$ at both sides of \eqref{wk_direction_inek} in the following
\begin{equation}
	\frac{1+\gamma_k^{\circ}}{\gamma_k^{\circ} } \frac{\lambda_k^{\circ}\beta_k}{\sigma_k^2} \bdg_k \bdg_k^H \bdw_k^{\circ} = \sum_{i=1}^K \frac{\lambda_i^{\circ} \beta_i}{\sigma_i^2} \bdg_i \bdg_i^H \bdw_k^{\circ} +  \bdw_k^{\circ} = \bdV^{\circ} \bdw_k^{\circ} \comma \label{wk_direction_identity}
\end{equation}
where $\bdV^{\circ} \in \Complex{M}{M}$ is only determined by $\{\lambda_k^{\circ}\}_{k=1}^K$ as follows
\begin{equation}
	\bdV^{\circ} = \sum_{k=1}^K \frac{\lambda_k^{\circ} \beta_k}{\sigma_k^2} \bdg_k \bdg_k^H + \bdI_M.
\end{equation} 
From \eqref{wk_direction_identity}, $\bdw_k^{\circ}$ can be written as
\begin{equation}
	\bdw_k^{\circ} = \Xinv{\bdV^{\circ}} \bdg_k \cdot \underbrace{ \frac{1+\gamma_k^{\circ}}{\gamma_k^{\circ} } \frac{\lambda_k^{\circ}\beta_k}{\sigma_k^2} \bdg_k^H \bdw_k^{\circ} }_{\text{scalar}}. \label{wk_Vinv_gk}
\end{equation}
By multiplying \eqref{wk_Vinv_gk} with $\bdg_k^H$ from the left side and then eliminating $\bdg_k^H \bdw_k^{\circ}$, we can derive
\begin{equation}
	\frac{\gamma_k^{\circ}}{1+\gamma_k^{\circ} } = \frac{\lambda_k^{\circ}\beta_k}{\sigma_k^2} \bdg_k^H \Xinv{\bdV^{\circ}} \bdg_k. \label{frac_one_add_gammak_gamma}
\end{equation}
From \eqref{frac_one_add_gammak_gamma}, it can be seen that $\gamma_k^{\circ}$ is also fully characterized by $\{\lambda_k^{\circ}\}_{k=1}^K$ as follows
\begin{equation}
	\gamma_k^{\circ} = \xinv{1-(\lambda_k^{\circ}\beta_k/\sigma_k^2) \bdg_k^H \Xinv{\bdV^{\circ}} \bdg_k} - 1. \label{gammak_lambda}
\end{equation}
Most importantly, from \eqref{wk_Vinv_gk}, we can see that the precoding vector $\bdw_k^{\circ}$ must be parallel to $\Xinv{\bdV^{\circ}} \bdg_k$ and thus, it can be expressed as
\begin{equation}
	\bdw_k^{\circ} = \sqrt{q_k^{\circ}} \cdot \frac{ \Xinv{\bdV^{\circ}} \bdg_k }{ \smallnorm{ \Xinv{\bdV^{\circ}} \bdg_k } }\comma \label{wk_qk_Vinv_gk}
\end{equation}
where $q_k^{\circ} = \smallnorm{\bdw_k^{\circ}}^2$ is the power of the precoding vector $\bdw_k^{\circ}$. By taking advantage of the condition in \eqref{grad_Lagrange_gammak_S2}, $q_k^{\circ}$ can be written as
\begin{equation}
	q_k^{\circ} = \frac{ \lambda_k^{\circ} \beta_k (\gamma_k^{\circ}+1) }{ \mu^{\circ} \sigma_k^2 } \smallnorm{ \Xinv{\bdV^{\circ}} \bdg_k }^2. \label{qk_mu}
\end{equation}
With the help of the power constraint $\sum_{k=1}^{K} \smallnorm{\bdw_k^{\circ}}^2 = \sum_{k=1}^{K} q_k^{\circ} = P$, the Lagrange multiplier $\mu^{\circ}$ can be expressed as
\begin{equation}
	\mu^{\circ} = \xinv{P} \sum_{k=1}^{K} \frac{ \lambda_k^{\circ} \beta_k (\gamma_k^{\circ}+1) }{ \sigma_k^2 } \smallnorm{ \Xinv{\bdV^{\circ}} \bdg_k }^2. \label{mu_Pinv_sum}
\end{equation}
From \eqref{qk_mu} and \eqref{mu_Pinv_sum}, it can be observed that $q_k^{\circ}$ is merely dependent on $\{\lambda_k^{\circ}\}_{k=1}^K$ as well. 
In summary, once the scalar variables $\{\lambda_k^{\circ}\}_{k=1}^K$ are known, the precoding vectors $\{\bdw_k^{\circ}\}_{k=1}^K$ can be computed by using \eqref{wk_qk_Vinv_gk} accordingly. 
After substituting $\gamma_k^{\circ}$ in \eqref{gammak_lambda} into $\log(1+\gamma_k)$, we can obtain that
\begin{align}
	\log(1+\gamma_k^{\circ}) &= - \log\left( 1-\frac{\lambda_k^{\circ}\beta_k}{\sigma_k^2} \bdg_k^H \Xinv{\bdV^{\circ}} \bdg_k \right) \notag \\
	&= - \log\det\left( \bdI_M -\frac{\lambda_k^{\circ}\beta_k}{\sigma_k^2} \bdg_k \bdg_k^H \Xinv{\bdV^{\circ}} \right) \notag \\
	&= \log \det \left( \sum_{i=1}^K \frac{\lambda_i^{\circ} \beta_i}{\sigma_i^2} \bdg_i \bdg_i^H + \bdI_M \right) \notag \\
	& \quad - \log\det\left( \sum_{i\ne k} \frac{\lambda_i^{\circ} \beta_i}{\sigma_i^2} \bdg_i \bdg_i^H + \bdI_M  \right)\comma
\end{align}
which is exactly equal to $r_k(\lambda_1^{\circ},\dots,\lambda_K^{\circ})$ in \eqref{rk_expression}. Next, we will show that the scalar variables $\{\lambda_k^{\circ}\}_{k=1}^K$ must satisfy $\sum_{k=1}^{K} \lambda_k^{\circ} = P$. Notice that the constraints in \eqref{Problem_precoding_gammak_S2} must hold with equality at the optimum, i.e.,
\begin{equation}
	\frac{\beta_k}{\gamma_k^{\circ} \sigma_k^2} \abs{\bdg_k^H \bdw_k^{\circ}}^2 = \sum_{i \ne k} \frac{\beta_k}{\sigma_k^2} \abs{\bdg_k^H \bdw_i^{\circ}}^2  + 1. \label{SINR_equal_gamma_optimum}
\end{equation}
Multiplying $\lambda_k^{\circ}$ at both sides of \eqref{SINR_equal_gamma_optimum} yields
\begin{equation}
	\frac{\lambda_k^{\circ}\beta_k}{\gamma_k^{\circ} \sigma_k^2} \abs{\bdg_k^H \bdw_k^{\circ}}^2 = \sum_{i \ne k} \frac{\lambda_k^{\circ}\beta_k}{\sigma_k^2} \abs{\bdg_k^H \bdw_i^{\circ}}^2  + \lambda_k^{\circ}. \label{Equality_SINR_lambda}
\end{equation}
On the other hand, we left-multiply \eqref{wk_direction_inek} with $\XH{\bdw_k^{\circ}}$ to obtain that
\begin{equation}
	\frac{\lambda_k^{\circ}\beta_k}{\gamma_k^{\circ} \sigma_k^2} \abs{ \bdg_k^H \bdw_k^{\circ} }^2 = \sum_{i \ne k} \frac{\lambda_i^{\circ} \beta_i}{\sigma_i^2} \abs{ \bdg_i^H \bdw_k^{\circ} }^2 + \norm{ \bdw_k^{\circ} }^2. \label{Equality_wk_direction_lambda}
\end{equation}
Consequently, we can derive that
\begin{align}
	\sum_{k=1}^K \lambda_k^{\circ} 
	&\stackeq{a} \sum_{k=1}^K \frac{\lambda_k^{\circ}\beta_k}{\gamma_k^{\circ} \sigma_k^2} \abs{\bdg_k^H \bdw_k^{\circ}}^2 - \sum_{k=1}^K \sum_{i \ne k} \frac{\lambda_k^{\circ}\beta_k}{\sigma_k^2} \abs{\bdg_k^H \bdw_i^{\circ}}^2 \notag \\
	&\stackeq{b} \sum_{k=1}^K \frac{\lambda_k^{\circ}\beta_k}{\gamma_k^{\circ} \sigma_k^2} \abs{ \bdg_k^H \bdw_k^{\circ} }^2 - \sum_{k=1}^K \sum_{i \ne k} \frac{\lambda_i^{\circ} \beta_i}{\sigma_i^2} \abs{ \bdg_i^H \bdw_k^{\circ} }^2 \notag \\
	&\stackeq{c} \sum_{k=1}^K \norm{ \bdw_k^{\circ} }^2 = P \comma
\end{align}
where (a) and (c) follow from \eqref{Equality_SINR_lambda} and \eqref{Equality_wk_direction_lambda}, respectively, and (b) comes from exchanging indices $k$ and $i$.
This concludes the proof.

\section{A minorizing function of $r_k$} 
\label[secinapp]{appendix_minorizing_rk_DL_instRx_proof}
We first consider a virtual UL multi-user single-input multiple-output (MU-SIMO) channel. 
In the virtual UL, each single-antenna UT transmits one data stream to a BS equipped with $M$ antennas. The received signal $\bdy \in \Complex{M}{1}$ at the BS can be written as 
\begin{equation}
	\bdy = \sum_{i=1}^{K} \sqrt{ \beta_i/\sigma_i^2 } \bdg_i \cdot \sqrt{\lambda_i }  d_i + \bdz\comma
\end{equation}
where $\sqrt{ \beta_i/\sigma_i^2 } \bdg_i$ is the channel vector between the UT $i$ and the BS, $\lambda_i \ge 0$ and $d_i$ are the transmit power and  data symbol of the UT $i$. The data symbol $d_i$ is assumed to have zero mean and unit variance, and $\bdz \sim \clCN(\bdzro,\bdI_M)$ is the additive complex Gaussian noise. 

We assume that the BS decodes the data streams of each UT without successive interference cancellation (SIC) \cite{Cover2006ElementsIT}. The BS uses a linear receiver $\bdu_k \in \Complex{M}{1}$ to recover the data symbol from UT $k$. Then, the recovered data symbol $\htd_k$ of UT $k$ can be written as 
\begin{align}
	\htd_k &= \bdu_k^H \bdy \notag \\
	&= \bdu_k^H \bdg_k \sqrt{ \frac{ \lambda_k \beta_k }{ \sigma_k^2 } } d_k + \sum_{i \ne k} \bdu_k^H \bdg_i \sqrt{ \frac{ \lambda_i \beta_i }{ \sigma_i^2 } } d_i + \bdu_k^H \bdz.
\end{align}
The virtual MSE (VMSE) of UT $k$ can be expressed as 
\begin{align}
	\VMSE_k &= \bbE\left\{ \smallabs{ \htd_k - d_k }^2 \right\} \notag \\
	&= \sum_{i=1}^{K} \abs{ \bdu_k^H \bdg_i }^2 \frac{ \lambda_i \beta_i }{ \sigma_i^2 } - 2\Re \left\{ \bdu_k^H \bdg_k \right\} \sqrt{ \frac{ \lambda_k \beta_k }{ \sigma_k^2 } } \notag \\
	&\qquad + \norm{ \bdu_k }^2 + 1. \label{VMSE_UTk_DL}
\end{align}
The $\bdu_k$ that minimizes $\VMSE_k$ is given by 
\begin{align}
	\bdu_k^{\vmmse} &= \arg \min_{\bdu_k} \VMSE_k \notag \\
	&= \Xinv{ \sum_{i=1}^{K} \frac{ \lambda_i \beta_i }{ \sigma_i^2 } \bdg_i \bdg_i^H + \bdI_M  } \bdg_k \sqrt{ \frac{ \lambda_k \beta_k }{ \sigma_k^2 } }\comma
\end{align}
and the corresponding virtual MMSE (VMMSE) of UT $k$ is given by
\begin{equation}
	\VMMSE_k = 1 - \frac{ \lambda_k \beta_k }{ \sigma_k^2 } \bdg_k^H \Xinv{\sum_{i=1}^{K} \frac{ \lambda_i \beta_i }{ \sigma_i^2 } \bdg_i \bdg_i^H + \bdI_M} \bdg_k.
\end{equation}
Then, $r_k$ can be rewritten as $r_k = -\log \VMMSE_k$.
Denote $\{\lambda_k^{(n)}\}_{k=1}^K$ as the scalar variables in the $n$th iteration. The VMMSE of UT $k$ in the $n$th iteration is given by
$\xiter{\VMMSE_k}{n} = \VMMSE_k\rvert_{\lambda_k=\lambda_k^{(n)},\forall k\in\clK}$.
By applying the concavity of $\log(\cdot)$, a minorizing function of $r_k$ can be derived as
\begin{align}
	r_k &= - \log\VMMSE_k \notag \\
	&\ge \xiter{r_k}{n} - \frac{ \VMMSE_k - \xiter{\VMMSE_k}{n}  }{ \xiter{\VMMSE_k}{n} } \notag \\
	&\stackgeq{a} \xiter{r_k}{n} + 1 - \frac{ \VMSE_k  }{ \xiter{\VMMSE_k}{n} } \triangleq h_k^{(n)}\comma \label{DL_rk_concavity}
\end{align}
where $\xiter{r_k}{n}$ is the computed $r_k$ in the $n$th iteration, (a) follows from the inequality $\VMMSE_k \le \VMSE_k$. Notice that $\VMSE_k$ is relevant to $\{ \lambda_k \}_{k=1}^K$ and $\bdu_k$. To make the last  inequality in \eqref{DL_rk_concavity} hold with equality at $\{\lambda_k^{(n)}\}_{k=1}^K$, we  choose $\bdu_k$ in $\VMSE_k$ as 
$\xiter{\bdu_k}{n} = \bdu_k^{\vmmse}\rvert_{\lambda_k=\lambda_k^{(n)},\forall k\in\clK}$.
Substituting $\xiter{\bdu_k}{n}$ into $\VMSE_k$ yields
\begin{equation}
	\frac{ \VMSE_k }{ \VMMSE_k^{(n)} } = \sum_{i=1}^{K}  \psi_{k,i}^{(n)} \frac{ \lambda_i \beta_i }{ \sigma_i^2 } - 2 \chi_{k}^{(n)} \sqrt{ \frac{ \lambda_k \beta_k }{ \sigma_k^2 } } + \delta_k^{(n)}\comma 
\end{equation}
where $\psi_{k,i}^{(n)}= \frac{ \smallabs{ \bdg_i^H \bdu_k^{(n)} }^2 }{ \VMMSE_k^{(n)} }$, $\chi_k^{(n)}= \frac{ \Re \{ \bdg_k^H \bdu_k^{(n)} \} }{ \VMMSE_k^{(n)} }$ and $\delta_k^{(n)}= \frac{ \smallnorm{ \bdu_k^{(n)} }^2 + 1 }{ \VMMSE_k^{(n)} }$.

%\ifCLASSOPTIONcaptionsoff
%\newpage
%\fi

\bibliographystyle{IEEEtran} 
\bibliography{IEEEabrv,satellite_integral_precoding_lib} 

\end{document}